\documentclass{article}

\usepackage[preprint,nonatbib]{neurips_2021}

\usepackage[utf8]{inputenc} 
\usepackage[T1]{fontenc}    
\usepackage{url}            
\usepackage{booktabs}       
\usepackage{amsfonts}       
\usepackage{nicefrac}       
\usepackage{microtype}      
\usepackage{xcolor}         

\usepackage{color,amsmath,amstext,amsfonts,amssymb, mathrsfs,mathtools}
\usepackage{graphicx,algorithmic,algorithm}
\usepackage{tikz}
\usepackage{setspace}
\usepackage{array} 
\usepackage{booktabs}  
\usepackage{bbm}
\usepackage{wasysym}
\usepackage{textcomp}
\usepackage{mathtools}
\usepackage{empheq}

\newtheorem{lemma}{Lemma}
\newtheorem{assumption}{Assumption}

\newtheorem{theorem}{Theorem}
\newtheorem{definition}{Definition}

\let\oldnl\nl
\newcommand{\nonl}{\renewcommand{\nl}{\let\nl\oldnl}}

\newcounter{l1}
\newcounter{l2}
\newcounter{l3}
\newcommand{\bdotlist}{\begin{list}{$\bullet$}{}}
\newcommand{\bboxlist}{\begin{list}{$\Box$}{}}
\newcommand{\bbboxlist}{\begin{list}{\raisebox{.005in}{{\tiny $\blacksquare$ \ \ }}}{}}
\newcommand{\bdashlist}{\begin{list}{$-$}{} }
\newcommand{\blist}{\begin{list}{}{} }
\newcommand{\barablist}{\begin{list}{\arabic{l1}}{\usecounter{l1}}}
\newcommand{\balphlist}{\begin{list}{(\alph{l2})}{\usecounter{l2}}}
\newcommand{\bAlphlist}{\begin{list}{\Alph{l2}.}{\usecounter{l2}}}
\newcommand{\bdiamlist}{\begin{list}{$\diamond$}{}}
\newcommand{\bromalist}{\begin{list}{(\roman{l3})}{\usecounter{l3}}}

\providecommand{\norm}[1]{\lVert#1\rVert}

\newcommand{\beq}{\begin{equation}}
\newcommand{\eeq}{\end{equation}}

\DeclarePairedDelimiter{\ceil}{\lceil}{\rceil}

\DeclarePairedDelimiterX{\Norm}[1]{\lVert}{\rVert}{#1}

\title{Online Learning Robust Control of Nonlinear Dynamical Systems}

\author{%
 Deepan Muthirayan \thanks{} \\
 University of California, Irvine\\
 Irvine, CA - 92697  \\
 \texttt{deepan.m@uci.edu} \\
 \And
 Pramod P. Khargonekar\thanks{} \\
 University of California, Irvine\\
 Irvine, CA - 92697  \\
 \texttt{pramod.khargonekar@uci.edu} \\
}

\begin{document}

\maketitle

\begin{abstract}
In this work we address the problem of  the online robust control of nonlinear dynamical systems perturbed by disturbance. We study the problem of attenuation of the total cost over a duration $T$ in response to the disturbances. The attenuation is defined as the ratio of the deviation of the total cost from the cost of the desired trajectory and the total energy in the disturbance. This is a harder problem than dynamic regret minimization because the comparator in the attenuation problem is zero total cost. We consider the setting where the cost function (at a particular time) is a general continuous function and adversarial, the disturbance is adversarial and bounded at any point of time. Our goal is to design a controller that can learn and adapt to achieve a certain level of attenuation. We analyse two cases (i) when the system is known and (ii) when the system is unknown. We measure the performance of the controller by the deviation of the controller's cost for a sequence of cost functions with respect to an attenuation $\gamma$, $R^p_t$. We propose an online controller and present guarantees for the metric $R^p_t$ when the maximum possible attenuation is given by $\overline{\gamma}$, which is a system constant. We show that when the controller has preview of the cost functions and the disturbances for a short duration of time and the system is known $R^p_T(\gamma) = O(1)$ when $\gamma \geq \gamma_c$, where $\gamma_c = \mathcal{O}(\overline{\gamma})$. We then show that when the system is unknown the proposed controller with a preview of the cost functions and the disturbances for a short horizon achieves $R^p_T(\gamma) = \mathcal{O}(N) + \mathcal{O}(1) + \mathcal{O}((T-N)g(N))$, when $\gamma \geq \gamma_c$, where $g(N)$ is the accuracy of a given nonlinear estimator and $N$ is the duration of the initial estimation period. We also characterize the lower bound on the required prediction horizon for these guarantees to hold in terms of the system constants.  
\end{abstract} 

\section{Introduction}

In this work we study the problem of regulating a dynamical system perturbed by disturbance. Restating \cite{goel2020regret}, the standard $\mathcal{H}_2$ and $\mathcal{H}_\infty$ control approaches for this problem are designed for a specific performance objective: (i) {\it the expected cost for random disturbance drawn from a specific distribution in the $\mathcal{H}_2$ case} and (ii) {\it the worst-case cost for adversarial disturbance in the $\mathcal{H}_{\infty}$ case}. These approaches can be poor if the disturbance belongs to a different class \cite{goel2020regret}. 

This naturally leads to the question whether the control can be dynamically adjusted to achieve desirable performance for any disturbance realization. Recently, this problem has been studied in online control from the point of view of regret minimization \cite{agarwal2019online, hazan2020nonstochastic, foster2020logarithmic}. In this framework, for a dynamically evolving system with state $x_t$, the online control policy $\pi = \{\pi_t\}_{1 \leq t \leq T}$ is chosen so as to minimize the regret  
\beq
R^\Pi_T = \left(\sum_{t=1}^T c_t(x_t, \pi_t(\{w_l\}_{1 \leq l\leq t}) - \inf_{\pi \in \Pi}\sum_{t=1}^T c_t(x_t, \pi(w))\right), \ w = \{w_t\}_{1 \leq t \leq T} \nonumber 
\eeq

In contrast, \cite{goel2020regret} approach this problem from the point of view minimizing the {\it dynamic regret}. The key difference is that the regret is defined against the best sequence of control actions in hindsight, not specific to a certain class of policies. The consideration of this regret is important from two perspectives: (i) the comparison is not restricted to a certain class of policies but is against the global optimal dynamic sequence of actions, (ii) as stated in \cite{goel2020regret} the online controller designed for this regret is more flexible, i.e., can respond to any type of disturbance, whereas linear feedback controllers can only be effective against either stochastic disturbance or adversarial disturbance. The dynamic regret is defined as follows:
\beq
R^d_T = \sum_{t=1}^T c_t(x_t, \pi_t(\{w_l\}_{1 \leq l\leq t}) - \inf_{\{u_t\}_{1 \leq t \leq T}}\sum_{t=1}^T c_t(x_t, u_t(w)), \ w = \{ w_t\}_{1 \leq t \leq T} \nonumber 
\eeq

In \cite{goel2020regret}, the authors derive the regret sub-optimal controller from the point of view of the dynamic regret for a known sequence of time-varying quadratic cost functions $c_t(.,.)$ for a known time-varying linear dynamical system. They show that for a specified attenuation $\gamma$ the regret sub-optimal controller satisfies
\begin{equation}
 R^d_T \leq \gamma \sum_{t=1}^T \norm{w_t}^2_2, \nonumber
\end{equation}
provided such a controller exists for the specified $\gamma$. While these settings study the performance with respect to a class of control policies, they do not provide guarantees for the actual performance: {\it the absolute attenuation of the cost of the system by the online adaptive controller in response to the disturbance}. The central problem we study in this work is {\it online guarantees for this problem of attenuation}. The attenuation is a harder notion of performance than dynamic regret because the bound on the attenuation gives a bound on the ratio of the dynamic regret and the total energy in the disturbance, defined as $\sum_{t=1}^T \norm{w_t}^2_2$, but not vice-versa. We study the setting where the controller may not have complete knowledge of the sequence of cost functions, and the system. Hence, the controller has to learn online to attenuate the cost of the system. The question is {\it how much the online adaptive controller can attenuate the total cost? At what rate does the total cost converge to this attenuation level with $T$?} To answer these questions, we study the deviation of the controller's cost for a sequence of cost functions, $\{c_t(.,.), t = 1,2,...\}$, of the state of the system $x_t$ and the control input $u_t$, with respect to an attenuation $\gamma$:
\begin{equation}
R^p_T(\gamma) = \left(\sum_{t=1}^T c_t(x_t, u_t) - \gamma \sum_{t=1}^T \norm{w_t}^2\right)_+. \nonumber 
\end{equation}
The metric as stated is a function of $\gamma$ because for the same sequence of control actions the deviation of the cumulative cost from the second term will depend on $\gamma$. We state the \textit{online robust problem} as follows:
\begin{equation}
\textit{1. What is the achievable} \ R^p_T(\gamma)? \hspace{2em} \textit{2. What is the smallest} \ \gamma \ \text{s.t.} \ R^p_T \ \text{is sub-linear?} \nonumber 
\end{equation}
The smallest $\gamma$ gives the maximum achievable {\it attenuation} and the corresponding $R^p_T(\gamma)$ quantifies the {\it convergence rate} to this attenuation: $R^p_T(\gamma)/T$. This is the central question we study in this work. The problem we study is a harder notion of performance than dynamic regret because the bound on the cost derived from $R^p_t$ is a bound on the dynamic regret and not vice versa.

In this work we present algorithms and guarantees for the questions posed above. We consider a setting similar to \cite{agarwal2019online}. The system the system is a general {\it non-linear dynamical system} perturbed by disturbance. The sequence of cost functions and the disturbances are adversarial and the disturbance at any time is bounded. We make minimal assumptions on the cost functions. We assume that they are continuous functions such that are lower bounded by $\underline{\alpha}\sigma(x)$, where $\sigma(.)$ is a general non-negative function. We assume that the desired trajectory is given by $\sigma(x) = 0$. Most distinctly, the cost functions need not be convex.

We present our guarantees for the achievable attenuation for the condition that the optimal cost-to-go function for the horizon $M$ at time $t$ is upper bounded by $\overline{\alpha}\sigma(x_t) + \overline{\gamma} \sum_{k=0}^{M-1} \norm{w_{t+k}}^2_2$. This assumption for the upper bound has two terms: (i) the first term is the contribution of the initial state to the optimal cost-to-go for the period $M$ from $t$ and (ii) the second term is the contribution of the disturbance over the period $M$. The factor $\overline{\gamma}$ is the maximum achievable attenuation. We note that the optimal cost-to-go for horizon $M$ for strongly convex quadratic cost functions for linear dynamical systems trivially satisfies such an upper bound. Hence, the setting we present is sufficiently general. This assumption does not in anyway restrict the problem and is a necessary assumption because the problem we address in this work has a solution only if a finite attenuation is achievable. 

\subsection{Contribution}

The online controller we propose and analyse is a Receding Horizon Controller (RHC) which has preview of the cost functions for a short horizon $M$ at time $t$. In a typical RHC setting, the controller optimizes the cost-to-go for a certain horizon and applies the first element of the computed optimal control sequence as the control input. This is then repeated at every time step. The RHC approach to control is a well established technique in control. The primary advantage of the RHC approach is that the optimization carried out at every step can incorporate any type of control input, system state and output constraints, and cost functions \cite{ mayne2000constrained, morari1999model, borrelli2017predictive}. The RHC controller we propose optimizes the cost-to-go for the short horizon $M$ to compute the control input for time $t$. It then applies the first element of the computed optimal control sequence for the horizon $M$ as the control input. This is repeated at every time step. We analyse the performance of RHC over a duration $T \gg M$. 

First, we present the setting where the system is known. We present: (i) the performance guarantee for the Receding Horizon Controller (RHC) when the controller has preview of the disturbance $w_k$ for the horizon $M$, i.e., for $t \leq k \leq t+M-1$, and (ii) present the performance guarantee when the controller does not have any preview of the disturbance. For the first case we show that RHC achieves $R^p_T(\gamma) = O(1)$ when $\gamma \geq \gamma_c$ and the horizon length $M > \overline{\alpha}^2/\underline{\alpha}^2+1$. We show that the achievable attenuation in this case is nearly optimal that is $\gamma_c = \mathcal{O}(\overline{\gamma})$. For the second case, when the preview of the disturbance is not available we guarantee that the overall cost is bounded by $\sim 2\overline{\gamma}MT\max_{w\in \mathcal{W}} \norm{w}^2_2 + \mathcal{O}(1)$. Hence, in this case we guarantee that achievable attenuation is nearly $\mathcal{O}(\overline{\gamma})$ of the maximum energy in the disturbance.

We then present the setting where the system is unknown and the disturbance preview is available. We present analysis for systems that are Bounded Input Bounded Stable (BIBO) and the cost functions are convex. We assume that the state transition function is parameterized and that only the parameter is unknown. 
We propose a generic online control framework that operates in two phases: the estimation phase followed by the control phase. The estimation phase runs for a period of $N$ time steps, at the end of which the online controller calculates an estimate $\hat{\theta}$ of $\theta$. In the control phase, the controller is the RHC that optimizes the cost-to-go for the estimated system. We present performance analysis for a black-box estimation procedure with accuracy given by: $\norm{\hat{\theta} -\theta}_2 \leq g(N)$, a bounded and decreasing function of $N$. We show that the overall online controller achieves $R^p_T(\gamma) = \mathcal{O}(N) + \mathcal{O}(1) + \mathcal{O}((T-N)g(N))$ when $\gamma \geq \gamma_c$ and the horizon length $M$ exceeds the same threshold as before. For linear systems, we trivially recover the results for the known system case since the estimation is trivial when the disturbance preview is available and the system is controllable. The framework we present provides a direct characterization to evaluate the various estimation approaches. 

\subsection{Notations}

For a sequence (of vectors/function) $\{a_{t}, 1 \leq t \leq T\}$, we denote $a_{1:t} = \{a_{\tau}, 1 \leq \tau \leq t\}$. For a matrix $M$, we denote its transpose by $M^\top$. We denote the maximum eigenvalue of a matrix $M$ by $\lambda_{\text{max}}(M)$. The two norm of a vector is denoted by $\norm{.}_2$. When two matrices $M_1$ and $M_2$ are related by $M_1 \geq M_2$ then it implies that $M_1 - M_2$ is positive semi-definite. Similarly, when $M_1 > M_2$ it implies that $M_1 - M_2$ is positive definite. We denote $\mathbb{R}^n$ as the $n-$dimensional euclidean space and $\mathbb{R}_{\geq 0}$ as the positive part of the real line. 

\section{Preliminaries}

We consider the control of a general non-linear dynamical system with disturbances in dynamics: 
\begin{equation}
x_{t+1} = f(x_t,u_t,w_t;\theta),
\label{eq:stateequation}
\end{equation} 
where $f$ is a continuous function, $x_{t} (\in \mathbb{R}^n$) is the system state, $u_t (\in \mathbb{R}^m)$ is the control input, $w_t (\in \mathbb{R}^n$) is the disturbance, the parameter $\theta$ could known or unknown. We make the assumption that the controller observes the full state.

We consider the setting where a fixed horizon preview of the future cost functions and disturbances are available to the algorithm at each time step $t$. In particular, the control input $u_{t}$ is computed as $u_{t} = \pi_{t}(x_{1:t}, u_{1:t-1},w_{1:t+M_{w}-1}, c_{1:t+M_{c}-1})$, where $M_{w}$ and $M_{c}$ are fixed horizon lengths and $\pi_{t}$ is the control policy at time $t$. The goal of the agent is to select a control policy $\pi = \pi_{1:T}$ in order to minimize the cumulative cost with respect to an attenuation $\gamma$. This can be formulated as the following optimization problem
\begin{align}
    \label{eq:basic-problem-1}
    & R^p_T(\gamma) = J(\pi; w_{1:T}, c_{1:T}) -\gamma \sum_{t=1}^T \norm{w_t}^2_2 :=   \sum^{T}_{t = 1} c_{t}(x_{t},u_{t}) -\gamma \sum_{t=1}^T \norm{w_t}^2_2 ,~~\text{where,} \nonumber \\
    & ~~u_{t} = \pi_{t}(x_{1:t}, u_{1:t-1},w_{1:t+M_w-1}, c_{1:t+M_c-1}). 
\end{align}
Our goal is to characterize the optimal $R^p_T(\gamma)$ and the maximum achievable attenuation.

The assumptions used throughout this paper are as follows:
\begin{assumption} (Disturbance)
{\it the disturbance $w_t \in \mathcal{W}$, where $\mathcal{W}$ is compact and $\norm{w_t}_2 \leq w_c$. The disturbance $w_t$ is adversarial for all $t$.}
\label{ass:noise}
\end{assumption}

\begin{assumption} (Cost Function)
{\it The function $c_t$ is Lipschitz continuous for all $t$, with a uniform Lipschitz constant $\alpha_c$ for all $t$ and there exists a constant $\underline{\alpha} > 0$ such that $c_t(x,u) \geq \underline{\alpha}\sigma(x) \geq 0$. The cost function $c_t(.,.)$ is adversarial for all $t$.}
\label{ass:stagecost}
\end{assumption}
The $\sigma(x)$ function could be any non-negative function. The assumption that the cost functions are Lipschitz continuous is a standard assumption \cite{li2019online, simchowitz2020improper}.

\section{Related Works}

The control of dynamical systems with uncertainties such as modeling errors, parametric uncertainty, and disturbances is a central challenge in control theory and so has been extensively studied. There is vast literature in the field of control on control synthesis for systems with such uncertainties. The {\it robust control} literature studies the problem of feedback control design for stability and performance guarantees with modeling uncertainty and disturbances; see \cite{skogestad2007multivariable}. The {\it adaptive control} literature studies the control of systems with parametric uncertainty; see \cite{sastry2011adaptive, ioannou2012robust, aastrom2013adaptive}.

{\it Online Control}: The field of online adaptive control is a widely studied topic. \cite{abbasi2011regret} studied the online Linear Quadratic Regulator (LQR) problem with unknown system and stochastic disturbances. 
The authors propose an adaptive algorithm that achieves $\sqrt{T}$ regret w.r.t the best linear control policy, which is the optimal policy. \cite{dean2018regret} propose an efficient algorithm for the same problem that achieves a regret of $\mathcal{O}(T^{2/3})$. \cite{cohen2019learning} and \cite{mania2019certainty} improved on this result by providing an efficient run-time algorithm with a regret guarantee of $\mathcal{O}(\sqrt{T})$ for the same problem. Mania et al. \cite{mania2019certainty} also establish $\mathcal{O}(\sqrt{T})$-regret for the partially observed Linear Quadratic Gaussian (LQG) setting. Recently, \cite{simchowitz2020naive} showed that $\mathcal{O}(\sqrt{T})$ is the optimal regret for the online LQR problem. \cite{cohen2018online} provide an $\mathcal{O}(\sqrt{T})$ algorithm for a variant of the online LQR where the system is known and noise is stochastic but the controller cost function is an adversarially chosen quadratic function.

\cite{agarwal2019online} consider the control of a known linear dynamic system with additive adversarial disturbance and an adversarial Lipschitz controller cost function. The controller they propose is a Disturbance Response Controller (DRC). They show that the the learning algorithm achieves $\mathcal{O}(\sqrt{T})$-regret with respect to the best DRC in hindsight. In a subsequent work \cite{agarwal2019logarithmic} show that a poly logarithmic regret is achievable for strongly convex controller cost and well conditioned stochastic disturbances. \cite{hazan2020nonstochastic} consider the same setting when the system is unknown and present an $\mathcal{O}(T^{2/3})$-regret algorithm. Recently, \cite{simchowitz2020improper} generalized these results to provide similar regret guarantees for the same setting with partial observation for both known and unknown systems.  

{\it Receding Horizon control}: Many receding horizon control based methods have been proposed for managing disturbances and uncertainties in the system dynamics. For example, some works handle disturbances or uncertainties by robust or chance constraints \cite{langson2004robust,goulart2006optimization,limon2010robust, tempo2012randomized, goulart2016robust}. Adaptive RHC techniques that adapt online when the system model is unknown have also been proposed \cite{fukushima2007adaptive, adetola2009adaptive, aswani2013provably,tanaskovic2019adaptive, bujarbaruah2019adaptive}. These methods primarily focus on constraint satisfaction, stability and in some cases performance improvement using the adapted models. In contrast to these works, we consider non-asymptotic performance of an online robust adaptive RHC. There are considerable amount of papers that provide performance analysis of RHC under both time-invariant costs \cite{angeli2011average, grune2014asymptotic, grune2015non} and time varying costs \cite{ferramosca2010economic, angeli2016theoretical, ferramosca2014economic, grune2017closed}. However, most of these studies focus on asymptotic performance. 

\section{Online Learning Robust Controller: Algorithms}

In this section, we present the online control algorithms for the setting where $M_c = M_w = M \ll T$. We present separate algorithms for the case where the system and the case where the system is unknown. The detailed discussion for the case where the disturbance is not accessible is deferred to the appendix. 

\subsection{System is known}

In this setting, we assume that the system dynamics is not parameterized by $\theta$. we assume that the algorithm has a fixed horizon preview of the future cost functions and disturbances. In particular, at each time $t$, algorithm has access to $c_{t:t+M-1}$ and $w_{t:t+M-1}$, in addition to the history of observation until $t$. We use a receding horizon control approach that minimizes the cost-to-go for a horizon $M$ with the previewed disturbances and cost functions. In particular, to find the control input $u_{t}$ at each time step $t$, the algorithm solves the following optimization problem:
\beq
\underset{\tilde{u}_{t:t+M-1}}{\inf} \sum_{j = t}^{t+M-1} c_j(\tilde{x}_j, \tilde{u}_j) \ \text{s.t.} \ \tilde{x}_{j+1} = f(\tilde{x}_j, \tilde{u}_j, w_j), \ \tilde{x}_t = x_t, \ \tilde{u}_j \in \mathcal{U}. 
\label{eq:cont-opt-distaccess}
\eeq
where $\mathcal{U}$ is a compact set. Given that $f,c_j$ are continuous functions the solution to the above optimization exists. Please see proof of Lemma \ref{lem:Lipschitz} for a detailed argument. We denote the solution of this optimization by $\text{MPC}^t(W_{t:t^e})$. The control input $u_t$ is set as the first element of the sequence $\text{MPC}^t(W_{t:t^e})$. We succinctly denote the control input by $\kappa_M(x_t)$. The overall online control algorithm is given in Algorithm \ref{alg:online-control-distaccess}.

\begin{algorithm}[]
\begin{algorithmic}[1]
\STATE \textbf{Input}: $M$
\FOR{t = 1,...,T}
\STATE Observe $x_t$, $\{c_j(.,.)\}_{t \leq j \leq t+M-1}$, $w_{t:t+M-1}$ 
\STATE Solve the optimiziation Eq. \eqref{eq:cont-opt-distaccess} to obtain $\tilde{u}_{t:t+M-1}$
\STATE Apply $\tilde{u}_t$ as the control input $u_t$
\ENDFOR
\end{algorithmic}
\caption{Online Learning Robust Control (Known System with Disturbance Preview)}
\label{alg:online-control-distaccess}
\end{algorithm}
 
\subsubsection{Discussion}

RHC algorithms are in general complex because of the explicit optimization that is carried out at every time step. The complexity could arise from the system models being too complicated to be directly used for prediction and sometimes the constraints in the problem. If the optimization is convex and the system is linear then the optimization can be solved efficiently via primal-dual interior point frameworks \cite{wright2019efficient}. Complex nonlinear RHC optimization such as Eq. \eqref{eq:cont-opt-distaccess} could be solved efficiently via online linearization of the nonlinear model. Such approaches have been shown to be empirically very effective for a wide range nonlinear control applications \cite{lawrynczuk2014computationally}. 

 
\subsection{Online Control Framework for Unknown System}

For the unknown system case, we assume the system dynamics is parametrized by an unknown parameter $\theta$, i.e.,
\begin{equation}
x_{t+1} = f(x_t,u_t,w_t;\theta). \nonumber 
\end{equation}
The online controller we propose operates in two phases: (i) {\it estimation phase}, and (ii) {\it control phase}. The estimation phase runs for a duration of $N$ time steps. We assume that the estimation algorithm is such that at the end of the estimation phase the estimated parameter $\hat{\theta}$ is such that $\norm{\hat{\theta}-\theta}_2 \leq g(N)$, a bounded and decreasing function of $N$. In the discussion below, we comment on the estimation procedures for specific instances of the system. In this work we present the performance analysis of the online controller without restricting the analysis to a specific estimation procedure. The performance guarantee is presented in terms of the accuracy of a given estimation algorithm $g(N)$. Thus, the analysis we present can be extended to any estimation procedure. 

In the control phase, the controller is the online receding horizon controller similar to the known system case. At each instant, the controller optimizes the cost-to-go for the horizon $M$ for the realized disturbance for the estimated system:
\beq
\underset{\tilde{u}_{t:t+M-1}}{\inf} \sum_{j = t}^{t+M-1} c_j(\tilde{x}_j, \tilde{u}_j) \ \text{s.t.} \ \tilde{x}_{j+1} = f(x_t,u_t,w_t;\hat{\theta}), \ \tilde{x}_{t} = x_t, \ \tilde{u}_j \in \mathcal{U}.
\label{eq:cont-opt-distaccess-unknown}
\eeq 
We later argue that the solution to this optimization exists. Please see proof of Lemma \ref{lem:Lipschitz} for a detailed argument. We denote the solution of this optimization by $\text{MPC}^t(w_{t:t+M-1};\hat{\theta})$. The control input $u_t$ is set as the first element of $\text{MPC}^t(w_{t:t+M-1};\hat{\theta})$. We succinctly denote the control input by $\kappa_M(x_t;\hat{\theta})$.

\subsubsection{Discussion}

In this work, we do not specify the estimation algorithm, rather we present a general framework and performance guarantee for a given black-box estimation procedure whose accuracy is characterized by $g(N)$, where $N$ is the duration of the estimation phase. 

The estimation procedure is trivially simple when the system is a linear controllable system. In this case, the system matrices can be estimated accurately in finite time by generating a sequence of control inputs that is persistently exciting over a finite interval (see \cite{moore1983persistence} for definition of persistence of excitation) and using least-squares to compute the estimate of the system matrices. In this case $N = O(1)$ and $g(N) = 0$. Hence, the online controller will be able to achieve the same performance as in the known system case. 

The estimation can also be simpler when the nonlinear system is linear in parameters \cite{lehrer2010parameter}, i.e., dynamics of the form $x_{k+1} = x_k + f(x_k,u_k) + g(x_k,u_k)\theta$. Under certain conditions on $g(x_k,u_k)$, an estimator can be designed to estimate $\theta$ in finite time \cite{lehrer2010parameter}. We note that a condition similar to persistence of excitation is required in this case too. 

Nonlinear systems are in general complex to estimate. A typical approach to estimate nonlinear systems is via the Koopman operator \cite{chen2018sample,mezic2020koopman}. The Koopman operator essentially specifies the transition in a lifted observable space. The Koopman operator can be approximately represented as a point in a linear vector space. The only disadvantage is that this representation requires a suitable set of basis functions to be known. Given this representation, linear estimation via least-squares can be applied to estimate the Koopman operator approximately \cite{chen2018sample}.  


The general version of the online control algorithm for the unknown system case is shown in Algorithm \ref{alg:online-control-distaccess}.

\begin{algorithm}[]
\begin{algorithmic}[1]
\STATE \textbf{Input}: $N$, $M$
\STATE Initialize: $\hat{\theta} = 0$
\FOR{t = 1,...,T}
\STATE Observe $x_t$, $\{c_j(.,.)\}_{t \leq j \leq t+M-1}$, $w_{t:t+m-1}$ 
\IF{$t < N$}
\STATE Apply control for estimation phase
\ELSE
\IF{$t == N$}
\STATE $\hat{\theta} \leftarrow$ estimation$(\mathcal{D})$, $\mathcal{D} = \{\{x_{k+1},x_k,u_k,w_k\}, {1 \leq k \leq N-1}\}$
\ENDIF
\STATE Solve the optimiziation Eq. \eqref{eq:cont-opt-distaccess-unknown} to obtain $\tilde{u}_{t:t+M-1}$
\STATE Apply $\tilde{u}_t$ as the control input $u_t$
\ENDIF
\ENDFOR
\end{algorithmic}
\caption{Online Learning Robust Control (Unknown System with Disturbance Preview)}
\label{alg:online-control-distaccess}
\end{algorithm}

\section{Analysis and Results}

In this section we present the performance analysis for the online controllers presented above. We first present the case where the system is known and then analyze the unknown case. 

\subsection{Analysis for Known System Case}

We define $V^t_M(x_t, w_{t:t+M-1})$ as the optimal value function of the optimization problem given in Eq. \ref{eq:cont-opt-distaccess}. Hence, $V^t_M(x_t, w_{t:t+M-1})$ is given by
\begin{align}
& V^t_M(x_t, w_{t:t+M-1}) = J^t_M(x_t, \tilde{u}_{t:t+M-1}, W_{t:t+M-1}), \nonumber \\
& J^t_M(x_t, \tilde{u}_{t:t+M-1}, w_{t:t+M-1}) =  \sum_{j = t}^{t+M-1} c_j(\tilde{x}_j, \tilde{u}_j) \ \text{s.t.} \ \tilde{x}_{j+1} = f(\tilde{x}_j, \tilde{u}_j, w_j),\  \tilde{x}_t = x_t. \nonumber
\end{align}

We note that this function exists and is continuous given that the solution to the optimization exists and is continuous (see Proof of Lemma \ref{lem:Lipschitz} for a more detailed argument). In the next lemma, we characterize the incremental change of the value of this function as the system evolves. We present this characterization for the class of systems that satisfies $V^t_M(x_t) \leq \overline{\alpha}\sigma(x_t) + \overline{\gamma} \sum_{k = t}^{t^e-1}\norm{w_k}^2_2$. This represents a sufficiently general class of systems. It includes the strongly convex quadratic cost functions for linear dynamical systems. As stated in the introduction, the first term in the bound represents the contribution to the cost by the initial state $x_t$ and the second term is the bound to the cost for the deviation effected by the disturbances. Here $\overline{\gamma}$ is the maximum achievable attenuation for this system.
\begin{lemma}
Suppose $t^e = t+M-1$, $w_{t:t^e}$ is accessible. Then there exists constants $\underline{\alpha}_V, \Gamma^{\gamma}_V > 0$ and $\Gamma_V$ such that for each $M \geq 2$ and $x_t \in \mathbb{R}^n$, for the closed-loop system $x_{t+1} = f(x_t, \kappa_M(x_t), w_t)$, $V^t_M(x_t,W_{t:t^e}) \geq \underline{\alpha}_V\sigma(x_t)$ and
\[
V^{t+1}_M(x_{t+1},W_{t+1:t^e+1}) - V^t_M(x_t,W_{t:t^e}) \leq \Gamma_V \sigma(x_t) + \Gamma^{\gamma}_V \sum_{k=t}^{t^e} \norm{w_k}^2_2,\]
where $\underline{\alpha}_V = \underline{\alpha}$, $\Gamma_V = \left(\frac{\overline{\alpha}^2}{\underline{\alpha}(M-1)}-\underline{\alpha}\right), \ \Gamma^{\gamma}_V = \overline{\gamma}\left(\frac{\overline{\alpha}}{\underline{\alpha}(M-1)}+1\right)$.
\label{lem:Lyapunov-existence}
\end{lemma}

Please see the Appendix for the proof. The result states that the change in the value of the optimal cost-to-go when the system evolves from $t$ to $t+1$ is bounded by (i) the cost of the initial state $x_t$ and (ii) some constant times the total energy of the disturbances for the duration $M$ starting from $t$. Next we use the above lemma to characterize an upper bound on the per step cost incurred by the online controller. 
\begin{lemma}
Under the conditions of Lemma \ref{lem:Lyapunov-existence}, suppose, $b = \overline{\gamma}\left(\frac{\underline{\alpha}}{\overline{\alpha}} + 1\right)$, $H \geq M$, $t_H = t+H$. Then, for $M > \overline{\alpha}^2/\underline{\alpha}^2 + 1$, and any $a$, s.t. $1 > a \geq 1 - \tilde{\epsilon}/\overline{\alpha}, 0 < \tilde{\epsilon} < \underline{\alpha} - \frac{\overline{\alpha}^2}{(M-1)\underline{\alpha}}$, the per-step cost of the closed-loop system $x_{t+1} = f(x_t, \kappa_M(x_t), w_t)$ at $t_H$ satisfies
\beq
c_{t+H}(x_{t_H},\kappa_M(x_{t_H})) \leq M_\lambda e^{-\lambda H}\sigma(x_t) + \sum_{j=t}^{t_H+M-2} M_{w,j} \norm{w_j}^2_2, \ \text{where}, \nonumber
\eeq
$M_\lambda = \overline{\alpha}, \ \lambda = -\ln{a}, \ M_{w,j} =  \left(ba^{H-M}/(1-a)+a^H\overline{\gamma}\right), \ \text{when} \ t \leq j \leq t^e, M_{w,j} = b/(1-a)a^{t_H-j-1}, \ \text{when} \ t^e+1 \leq j \leq t_H-1, \ M_{w,j} = b/(1-a), \text{when} \ t_H \leq j \leq t_H+M-2$.
\label{lem:cost-bound}
\end{lemma}

Please see the Appendix for the proof. The results imply that the cost at a time $t+H$ is bounded by two terms: (i) the contribution of the cost from the initial state $x_t$ that is exponentially decaying with $H$ and (ii) the contribution from the disturbances seen by the system so far. From the form of the coefficients $M_{w,j}$ it is clear that the online controller exponentially diminishes the contribution of the disturbances from earlier times. In the next theorem we characterize the performance of the online controller. 
\begin{theorem}
Under the conditions of Lemma \ref{lem:Lyapunov-existence}, Then, for $T \geq M+1$, $M > \overline{\alpha}^2/\underline{\alpha}^2 + 1$, and any $a$ s.t. $1 > a \geq 1 - \tilde{\epsilon}/\overline{\alpha}, 0 < \tilde{\epsilon} < \underline{\alpha} - \frac{\overline{\alpha}^2}{(M-1)\underline{\alpha}}$, the total cost incurred by the closed-loop system $x_{t+1} = f(x_t, \kappa_M(x_t), w_t)$ satisfies
\begin{equation}
\sum_{t=1}^{T} c_{t}(x_{t},\kappa_M(x_{t})) - \gamma \sum_{t=1}^T \norm{w_t}^2_2 \leq \mathcal{O}(1), \ \gamma \geq \gamma_c \nonumber
\end{equation}
where $\gamma_c = \frac{b\left((M-1)(1-a)+1\right)}{(1-a)^2}$.
\label{thm:cost-regret}
\end{theorem}

Please see the Appendix for the proof. It follows from the result that the achievable attenuation by the RHC approach is $\gamma_c$. 
 Given the bounds on $a$ we can choose $a$ and $M$ such that $2\left(1+\frac{\underline{\alpha}}{\overline{\alpha}}\right)\left(\frac{\tilde{\gamma}\overline{\alpha}^2}{\underline{\alpha}^2} +\tilde{\gamma}^2\right) \overline{\gamma} \gtrapprox \gamma_c \geq 2\overline{\gamma}$, where $\tilde{\gamma} = \frac{\ceil{\frac{\overline{\alpha}^2}{\underline{\alpha}^2}}\underline{\alpha}}{\overline{\alpha}\left(\frac{\underline{\alpha}^2}{\overline{\alpha}^2}\ceil{\frac{\overline{\alpha}^2}{\underline{\alpha}^2}}-1\right)}$. This means that the RHC approach cannot beat $2\overline{\gamma}$ and that $\gamma_c = \mathcal{O}(\overline{\gamma})$.

\subsection{Analysis for Unknown System Case}

Here, we present the performance analysis for a black-box estimation procedure with accuracy $\norm{\hat{\theta}-\theta}_F \leq g(N)$. Let
\begin{align}
& V^t_M(x_t, w_{t:t+M-1};\hat{\theta}) = J^t_M(x_t, \tilde{u}_{t:t+M-1}, w_{t:t+M-1};\hat{\theta}), \nonumber \\
& J^t_M(x_t, \tilde{u}_{t:t+M-1}, w_{t:t+M-1}) =  \sum_{j = t}^{t+M-1} c_j(\tilde{x}_j, \tilde{u}_j) \ \text{s.t.} \ \tilde{x}_{j+1} = f(\tilde{x}_j, \tilde{u}_j, w_j; \hat{\theta}),\  \tilde{x}_t = x_t. \nonumber
\end{align}

Let $U$ and $\tilde{U}$ be two control sequence of length $M$. Define
\begin{equation}
J^t_{1,M}(x_t,U,\tilde{U},w_{t:t^e};\theta) = J^t_M(x_t,U,W_{t:t^e};\theta)- J^t_M(x_t,\tilde{U},w_{t:t^e};\theta). \nonumber 
\end{equation}

We argue in Lemma \ref{lem:Lipschitz} that the solution to the RHC optimization $\text{MPC}^t(w_{t:t^e};\hat{\theta})$ exists. Hence the function $V^t_M(x_t, w_{t:t^e};\hat{\theta})$ is well defined.
\begin{definition}
We say that a function $f_1:\mathbb{R}^p\times \mathbb{R}^p \times \mathbb{R}^q \rightarrow \mathbb{R}$, $f_1 \in \mathcal{L}^{\alpha,\beta}$ provided $f_1(s,\tilde{s},\lambda) - f_1(s,\tilde{s},\tilde{\lambda}) \leq c\norm{s-\tilde{s}}^\alpha_2\norm{\lambda-\tilde{\lambda}}^\beta_2$.
\end{definition}

We make the following additional assumptions on the system and the cost functions. These assumptions are required to establish Lipschitz continuity of the solution $\text{MPC}^t(w_{t:t^e};\hat{\theta})$. This is the key additional step that is required to extend the analysis technique applied to the known system case.
\begin{assumption} (System and Cost)
(i) The cost is time invariant, (ii) The system given by $x_{t+1} = f(x_t,u_t,w_t;\theta)$ is Bounded Input Bounded Output (BIBO) stable; (iii) the condition $\sigma(x) = 0$ is bounded trajectory; (iv) the estimation accuracy $g(N)$ can be guaranteed by bounded excitation; (v) $\theta \in \Theta$, a compact set and $\norm{\theta}_2 \leq S$, and $f$ is continuous for all $\theta \in \Theta$; (vi) the function $f(x_t,u_t,w_t;\theta)$ is Lipschitz, i.e., $f(x'_t,u'_t,w_t;\theta) - f(x_t,u_t,w_t;\theta) \leq \alpha_f\norm{\theta}_2\left(\norm{x'_t-x_t}_2 + \norm{u'_t-u_t}_2\right)$ (trivially includes linear systems); (vii) the cost function $c_t(.,.)$ is convex in both arguments and strongly convex in the second argument (includes LQR); (viii) $J^t_{1,M} = f_1 \in \mathcal{L}^{1,1}$ wherein $s = U, \tilde{s} = \tilde{U}, \lambda = \theta, \tilde{\lambda} = \tilde{\theta}$; (ix) $\hat{\theta} \in \Theta$.
\label{ass:sys}
\end{assumption}

The assumption that the estimation accuracy $g(N)$ can be guaranteed by bounded excitation is trivially satisfiable for linear controllable systems. The assumption that the system is BIBO is required to ensure that the total cost incurred during the estimation phase scales only linearly with $N$.
\begin{lemma}
Under Assumption \ref{ass:sys} and the cost function is time invariant, it holds that
\begin{equation}
V^t_M(.,.;\hat{\theta}) - V^t_M(.,.;\theta) \leq \alpha_V\norm{\hat{\theta}-\theta}_2, \ \kappa_M(x_t;\hat{\theta}) - \kappa_M(x_t;\theta) \leq \alpha_\kappa \norm{\hat{\theta}-\theta}_2. \nonumber 
\end{equation}
\label{lem:Lipschitz}
\end{lemma}

To establish this result we draw from results on Lipschitz continuity of solutions to parametric optimization \cite{quincampoix2008parameterized}. Please see the Appendix for the detailed proof. In the next lemma, we characterize the incremental change of the value of the function $V^t_M(.,.;\hat{\theta})$ in the control phase as the closed loop system evolves from $t$ to $t+1$. We establish these results for the same class of systems assumed for the known system case, i.e., systems for which $V^t_M(x_t, W_{t:t^e};\theta) \leq \overline{\alpha}\sigma(x_t) + \overline{\gamma} \sum_{k=t}^{t^e-1} \norm{w_k}^2_2$ and additionally Assumption \ref{ass:sys}.
\begin{lemma}
Suppose Assumptions \ref{ass:noise}, \ref{ass:stagecost} \ref{ass:sys} hold. Suppose $t^e = t_M-1$, $W_{t:t^e}$ is accessible. Then there exists constants $\underline{\alpha}_V, \Gamma^{\gamma}_V > 0$ and $\Gamma_V$ such that for each $M \geq 2$ and $x_t \in \mathbb{R}^n$, for the closed-loop system $x_{t+1} = f(x_t, \kappa_M(x_t;\hat{\theta}), w_t)$ in the control phase satisfies $V^t_M(x_t,W_{t:t^e};\hat{\theta}) \geq \underline{\alpha}_V\sigma(x_t)$ and
\[
V^{t+1}_M(x_{t+1},W_{t+1:t^e+1};\hat{\theta}) - V^t_M(x_t,W_{t:t^e};\hat{\theta}) \leq \Gamma_V \sigma(x_t) + \Gamma^{\gamma}_V \sum_{k=t}^{t^e} \norm{w_k}^2_2 +\Gamma^\theta_V \norm{\hat{\theta}-\theta}_2,\]
where $\Gamma^\theta_V = \left(2\alpha_V+ \alpha_c\tilde{\alpha}_f \alpha_\kappa(M-1)\left(\frac{\overline{\alpha}}{\underline{\alpha}(M-1)}+1\right)\right)$, $\underline{\alpha}_V, \Gamma_V, \Gamma^{\gamma}_V$ are the same as in Lemma \ref{lem:Lyapunov-existence} and $\tilde{\alpha}_f = \max_{0\leq k \leq M-2} \alpha^{k+1}_fS^{k+1}$. 
\label{lem:Lyapunov-existence-unknownsystem}
\end{lemma}

Please see the Appendix for the Proof. We note that the first two terms in the bound are the same as in the known system case. The last term arises from the error in the parameter estimate $\hat{\theta}$. Next we use the above lemma to characterize an upper bound on the per step cost incurred by the online controller in the control phase.  
\begin{lemma}
Under the conditions of Lemma \ref{lem:Lyapunov-existence-unknownsystem}, suppose $b = \overline{\gamma}\left(\frac{\underline{\alpha}}{\overline{\alpha}} + 1\right)$, $H \geq M$, $t_H = t+H$. Then,  $M > \overline{\alpha}^2/\underline{\alpha}^2 + 1$, and any $a$, s.t. $1 > a \geq 1 - \tilde{\epsilon}/\overline{\alpha}, 0 < \tilde{\epsilon} < \underline{\alpha} - \frac{\overline{\alpha}^2}{(M-1)\underline{\alpha}}$, the per-step cost for the closed-loop system $x_{t+1} = f(x_t, \kappa_M(x_t;\hat{\theta}), w_t)$ at $t_H$ in the control phase satisfies
\beq
c_{t+H}(x_{t_H},\kappa_M(x_{t_H};\hat{\theta})) \leq M_\lambda e^{-\lambda H}\sigma(x_t) + \sum_{j=t}^{t_H+M-2} M_{w,j} \norm{w_j}^2_2 + M_{\theta}\norm{\hat{\theta}-\theta}_2, \ \text{where}, \nonumber
\eeq
$M_\lambda = \overline{\alpha}, \ \lambda = -\ln{a}, \ M_{w,j} =  \left(ba^{H-M}/(1-a)+a^H\overline{\gamma}\right), \ \text{when} \ t \leq j \leq t^e, M_{w,j} = \frac{b}/(1-a)a^{t_H-j-1}, \ \text{when} \ t^e+1 \leq j \leq t_H-1, \ M_{w,j} = b/(1-a), \ \text{when} \ t_H \leq j \leq t_H+M-2$, $M_\theta = \left( c(1-a^H)(1-a) +a^H \alpha_V\right)$, where $c = \left(\alpha_V\left(2+\frac{\tilde{\epsilon}}{\overline{\alpha}}\right)+ \alpha_c\tilde{\alpha}_f(M-1)\left(\frac{\underline{\alpha}-\tilde{\epsilon}}{\overline{\alpha}} + 1\right)\right)$.
\label{lem:cost-bound-unknownsystem}
\end{lemma}

Please see the Appendix for the proof. We note that the per-step cost in the control phase in this case is bounded by three terms; the first two terms are the same as before while the additional third term arises from the error in the parameter estimate $\hat{\theta}$. In the next theorem we characterize the performance of the online controller. 
\begin{theorem}
Suppose the conditions of Lemma \ref{lem:Lyapunov-existence-unknownsystem}. Then, for $T \geq M+1$, $M > \overline{\alpha}^2/\underline{\alpha}^2 + 1$, and any $a$ s.t. $1 > a \geq 1 - \tilde{\epsilon}/\overline{\alpha}, 0 < \tilde{\epsilon} < \underline{\alpha} - \frac{\overline{\alpha}^2}{(M-1)\underline{\alpha}}$, the total cost incurred by the closed-loop system $x_{t+1} = f(x_t, \kappa_M(x_t;\hat{\theta}), w_t)$ satisfies
\begin{equation}
\sum_{t=1}^{T} c_{t}(x_{t},\kappa_M(x_{t})) - \gamma \sum_{t=1}^T \norm{w_t}^2_2 \leq \mathcal{O}(N) + \mathcal{O}(1) + \mathcal{O}((T-N+1)g(N)), \ \gamma \geq \gamma_c \nonumber
\end{equation}
where $\gamma_c = \frac{b\left((M-1)(1-a)+1\right)}{(1-a)^2}$.
\label{thm:cost-regret-unknownsystem}
\end{theorem}
{\em Proof}: By Assumption \ref{ass:sys} that the system is BIBO and the accuracy $g(N)$ can be guaranteed by bounded excitation it follows that the cost for the estimation phase is bounded by $\gamma \sum_{t=1}^{N-1} \norm{w_t}^2_2 + \mathcal{O}(N)$. Then from Lemma \ref{lem:cost-bound-unknownsystem}, the fact that $\norm{\hat{\theta} -\theta}_2 \leq g(N)$, following steps similar to proof of Theorem \ref{thm:cost-regret} it follows that the bound for the cost for the control phase is $\gamma \sum_{t=N}^{T} \norm{w_t}^2_2 + \mathcal{O}(1) + \mathcal{O}((T-N+1)g(N))$. Combining the two observations the result follows $\blacksquare$

The result presented characterizes the achievable performance by the RHC approach for a given black-box estimation procedure. If $N = \mathcal{O}(1), g(N)=0$, as is the case with linear controllable systems, we recover the results for the known system case, where the RHC approach achieves the attenuation $\gamma_c$. Similarly, for nonlinear systems linear in unknown parameters, as argued earlier, $N = \mathcal{O}(1), g(N)=0$, and so the RHC approach will achieve the attenuation $\gamma_c$ in this case too. For a general estimation procedure for which $g(N) = \mathcal{O}(1/\sqrt{N})$, the optimal duration for the estimation phase is $N = \mathcal{O}(T^{2/3})$ and $R^p_T \leq \mathcal{O}(T^{2/3})$.  

\bibliographystyle{plain}
\bibliography{Refs.bib}

\section*{Appendix A: Proof of Lemma \ref{lem:Lyapunov-existence}}

We note that
\begin{equation}
V^t_M(x_t,W_{t:t^e}) \geq c_t(x_t, \kappa_M(x_t)) \geq \underline{\alpha}\sigma(x_t). \nonumber
\end{equation}
The last inequality follows directly from Assumption \ref{ass:stagecost}. Hence, $\underline{\alpha}_V = \underline{\alpha} > 0$. We introduce some notation for simplifying the presentation of the analysis. If $\text{MPC}^t(W_{t:t^e}) = \{\tilde{u}^*_t, \tilde{u}^*_{t+1},...,\tilde{u}^*_{t^e}\}$, then let $\text{MPC}^t( W_{t:t^e})[i:j] = \{\tilde{u}^*_{t+i-1},..., \tilde{u}^*_{t+j-1}\}, \ i \leq j$. We denote $\text{MPC}^t(W_{t:t^e})$ by $\text{M}^t_p$, and $W_{t:t^e}$ by $W_{t}$. 

Let $\phi^t(k, x_t, \text{M}^t_p, W_t)$ denote the state the system evolves to at time $t+k$ starting at $x_t$, under the control sequence $\text{M}^t_p[1:k]$ and disturbance $W_{t}$. From the definition of $V^{t}_M(x_{t},W_t)$ it follows that
\begin{align}
& V^{t+1}_M(x_{t+1},W_{t+1}) - V^{t}_M(x_{t},W_{t}) = J^{t+1}_M(x_{t+1}, \text{M}^{t+1}_p, W_{t+1}) - J^{t}_M(x_t, \text{M}^t_p, W_{t}) \nonumber \\
& = \sum_{k = 0}^{M-1} c_{t+1+k}(\phi^{t+1}(k, x_{t+1}, \text{M}^{t+1}_p, W_{t+1}), \text{M}^{t+1}_p[k+1]) - \sum_{k = 0}^{M-1} c_{t+k}(\phi^t(k, x_t, \text{M}^t_p, W_t), \text{M}^t_p[k+1]). \nonumber
\end{align}

The second step just follows from the definition of $J^{t}_M(.,.,.)$. Hence,
\begin{align}
& V^{t+1}_M(x_{t+1},W_{t+1}) - V^{t}_M(x_{t},W_{t})  = \sum_{k = 0}^{j} c_{t+1+k}(\phi^{t+1}(k, x_{t+1}, \text{M}^{t+1}_p, W_{t+1}), \text{M}^{t+1}_p[k+1]) \nonumber \\
& + \sum_{k = j+1}^{M-1}c_{t+1+k}(\phi^{t+1}(k, x_{t+1}, \text{M}^{t+1}_p, W_{t+1}), \text{M}^{t+1}_p[k+1]) - \sum_{k = 0}^{M-1} c_{t+k}(\phi^t(k, x_t, \text{M}^t_p, W_t), \text{M}^t_p[k+1]) \nonumber \\
& \leq \sum_{k = 0}^{j-1} c_{t+1+k}(\phi^{t+1}(k, x_{t+1}, \tilde{\text{M}}^{t+1}_p, W_{t+1}), \tilde{\text{M}}^{t+1}_p[k+1]) \nonumber \\
& + \sum_{k = j}^{M-1}c_{t+1+k}(\phi^{t+1}(k, x_{t+1}, \tilde{\text{M}}^{t+1}_p, W_{t+1}), \tilde{\text{M}}^{t+1}_p[k+1]) - \sum_{k = 0}^{M-1} c_{t+k}(\phi^t(k, x_t, \text{M}^t_p, W_t), \text{M}^t_p[k+1]), \nonumber
\end{align}

where $\tilde{\text{M}}^{t+1}_p = [\text{M}^{t}_p[2:j+1], \text{M}^{*j,t+1}_p[j+1:M]]$, where $\text{M}^{*j,t+1}_p[j+1:M]$ is the optimal control sequence given the control for the first $j$ time steps is $\text{M}^{t}_p[2:j+1]$. The second inequality follows from the fact that $\tilde{\text{M}}^{t+1}_p$ is sub-optimal compared to $\text{M}^{t+1}_p$. Now by definition
\begin{align}
& \sum_{k = 0}^{M-1} c_{t+k}(\phi^t(k, x_t, \text{M}^t_p, W_t), \text{M}^t_p[k+1]) = c_t(x_t,\text{M}^t_p[1])  \nonumber \\
& + \sum_{k = 0}^{M-2} c_{t+1+k}(\phi^{t+1}(k, x_{t+1}, \text{M}^t_p[2:M], W_{t+1}), \text{M}^t_p[k+2]). \nonumber  
\end{align}

Substituting this in the above inequality, the first term on the right, the summation from $k=0$ to $j$ gets cancelled. Hence, we have
\begin{align}
& V^{t+1}_M(x_{t+1},W_{t+1}) - V^{t}_M(x_{t},W_{t}) \leq \sum_{k = j}^{M-1}c_{t+1+k}(\phi^{t+1}(k, x_{t+1}, \tilde{\text{M}}^{t+1}_p, W_{t+1}), \tilde{\text{M}}^{t+1}_p[k+1]) \nonumber \\
& - c_t(x_t,\text{M}^t_p[1]). \nonumber 
\end{align}

By definition of $\tilde{\text{M}}^{t+1}_p$, the first sum on the right is the optimal cost-to-go starting from the state $\phi^{t+1}(j, x_{t+1}, \tilde{\text{M}}^{t+1}_p, W_{t+1})$ for the horizon $M-j$ from $t+j$. Hence, 
\begin{equation}
V^{t+1}_M(x_{t+1},W_{t+1}) - V^{t}_M(x_{t},W_{t}) \leq V^{t+j}_{M-j}(\phi^{t+1}(j, x_{t+1}, \tilde{\text{M}}^{t+1}_p, W_{t+1}), W_{t+j}) - c_t(x_t,\text{M}^t_p[1]). \nonumber 
\end{equation}

Once again by the assumption in the Lemma we have that
\begin{equation}
V^{t+j}_{M-j}(\phi^{t+1}(j, x_{t+1}, \tilde{\text{M}}^{t+1}_p, W_{t+1}), W_{t+j}) \leq \overline{\alpha} \sigma(\phi^{t+1}(j, x_{t+1}, \tilde{\text{M}}^{t+1}_p, W_{t+1})) + \overline{\gamma} \sum_{k=t+j}^{t+M-2} \norm{w_k}^2_2. \nonumber
\end{equation}

Hence, 
\begin{align}
& V^{t+1}_M(x_{t+1},W_{t+1}) - V^{t}_M(x_{t},W_{t}) \leq \overline{\alpha} \sigma(\phi^{t+1}(j, x_{t+1}, \tilde{\text{M}}^{t+1}_p, W_{t+1})) \nonumber \\
& + \overline{\gamma} \sum_{k=t+j}^{t+M-2} \norm{w_k}^2_2 - c_t(x_t,\text{M}^t_p[1]). \nonumber 
\end{align}

Now, $\phi^{t+1}(j, x_{t+1}, \tilde{\text{M}}^{t+1}_p, W_{t+1}) = \phi^{t+1}(j, x_{t+1}, \text{M}^{t}_p[2:j+1], W_{t+1})$. Hence,
\begin{align}
& V^{t+1}_M(x_{t+1},W_{t+1}) - V^{t}_M(x_{t},W_{t}) \leq \overline{\alpha} \sigma(\phi^{t+1}(j, x_{t+1}, \text{M}^{t}_p[2:j+1], W_{t+1})) \nonumber \\
& + \overline{\gamma} \sum_{k=t+j}^{t+M-2} \norm{w_k}^2_2 - c_t(x_t,\text{M}^t_p[1]).
\label{eq:lemma1-Vdiff-ub}
\end{align}

From Assumption \ref{ass:stagecost} and the fact that $V^t_M(x_t, W_{t:t^e}) \leq \overline{\alpha}\sigma(x_t) + \overline{\gamma} \sum_{k=t}^{t^e-1} \norm{w_t}^2_2$ for any $t$ it follows that
\begin{align}
& \underline{\alpha} \sum_{k = 0}^{M-2} \sigma(\phi^{t+1}(k, x_{t+1}, \text{M}^{t}_p[2:k+1], W_{t+1})) \leq \sum_{k = 0}^{M-2} c_{t+1+k}(\phi^{t+1}(k, x_{t+1}, \text{M}^{t}_p[2:M], W_{t+1}), \text{M}^{t}_p[k+2]) \nonumber \\
& \leq c_t(x_t,\text{M}^{t}_p[1])+ \sum_{k = 0}^{M-2} c_{t+1+k}(\phi^{t+1}(k, x_{t+1}, \text{M}^{t}_p[2:M], W_{t+1}), \text{M}^{t}_p[k+2]) \leq \overline{\alpha}\sigma(x_{t}) + \overline{\gamma} \sum_{k=t}^{t^e} \norm{w_k}^2_2. \nonumber 
\end{align}

The above inequality implies that $\exists \ j^{*} \in \{0,1,...,M-2\}$ such that
\begin{equation}
\sigma(\phi^{t+1}(j^{*}, x_{t+1}, \text{M}^{t}_p[2:j^{*}+1], W_{t+1})) \leq \frac{1}{\underline{\alpha}(M-1)}\left(\overline{\alpha}\sigma(x_{t}) + \overline{\gamma} \sum_{k=t}^{t^e} \norm{w_k}^2_2\right). \nonumber 
\end{equation}

Then using this in Eq. \eqref{eq:lemma1-Vdiff-ub} it we get that
\begin{align}
& V^{t+1}_M(x_{t+1},W_{t+1}) - V^{t}_M(x_{t},W_{t}) \leq \overline{\alpha} \sigma(\phi^{t+1}(j^{*}, x_{t+1}, \text{M}^{t}_p[2:j^{*}+1], W_{t+1})) \nonumber \\
& + \overline{\gamma} \sum_{k=t+j^{*}}^{t+M-2} \norm{w_k}^2_2 - c_t(x_t,\text{M}^t_p[1]) \nonumber \\
& \leq \frac{\overline{\alpha}}{\underline{\alpha}(M-1)}\left(\overline{\alpha}\sigma(x_{t}) + \overline{\gamma} \sum_{k=t}^{t^e} \norm{w_k}^2_2\right) + \overline{\gamma} \sum_{k=t+j^{*}}^{t+M-2} \norm{w_k}^2_2 - c_t(x_t,\text{M}^t_p[1]) \nonumber\\
& = \frac{\overline{\alpha}^2}{\underline{\alpha}(M-1)}\sigma(x_{t}) - c_t(x_t,\text{M}^t_p[1]) + \overline{\gamma}\left(\frac{\overline{\alpha}}{\underline{\alpha}(M-1)}+1\right)\left(\sum_{k=t}^{t^e} \norm{w_k}^2_2\right). \nonumber
\end{align}

Then using the fact that $c_t(x_t,\text{M}^t_p[1]) \geq \underline{\alpha}\sigma(x_t)$ we get that
\begin{align}
& V^{t+1}_M(x_{t+1},W_{t+1}) - V^{t}_M(x_{t},W_{t}) \leq \frac{\overline{\alpha}^2}{\underline{\alpha}(M-1)}\sigma(x_{t}) - \underline{\alpha}\sigma(x_t) + \overline{\gamma}\left(\frac{\overline{\alpha}}{\underline{\alpha}(M-1)}+1\right)\left(\sum_{k=t}^{t^e} \norm{w_k}^2_2\right) \nonumber \\
& = \left(\frac{\overline{\alpha}^2}{\underline{\alpha}(M-1)}-\underline{\alpha}\right)\sigma(x_{t}) + \overline{\gamma}\left(\frac{\overline{\alpha}}{\underline{\alpha}(M-1)}+1\right)\left(\sum_{k=t}^{t^e} \norm{w_k}^2_2\right). \nonumber
\end{align}

Hence, 
\begin{equation}
\Gamma_V = \left(\frac{\overline{\alpha}^2}{\underline{\alpha}(M-1)}-\underline{\alpha}\right), \ \Gamma^{\gamma}_V = \overline{\gamma}\left(\frac{\overline{\alpha}}{\underline{\alpha}(M-1)}+1\right). \nonumber
\end{equation}

$\blacksquare$ 

\section*{Appendix B: Proof of Lemma \ref{lem:cost-bound}}

Since $M > \overline{\alpha}^2/\underline{\alpha}^2+1$, there exists $\tilde{\epsilon} \in (0,\underline{\alpha})$ such that 
\begin{equation}
M > \overline{\alpha}^2/\left(\underline{\alpha}(\underline{\alpha}-\tilde{\epsilon})\right) + 1 \geq \overline{\alpha}/\left(\underline{\alpha}-\tilde{\epsilon}\right)+1. \nonumber
\end{equation}

For convenience of presentation in the following we denote $V^{t}_M(x_{t},W_{t:t^e})$ by $V^{t}_M$. From Lemma \ref{lem:Lyapunov-existence} it follows that
\begin{equation}
V^{t+1}_M - V^{t}_M \leq \Gamma_V \sigma(x) + \Gamma^{\gamma}_V \sum_{k=t}^{t^e} \norm{w_k}^2_2, \ \Gamma_V = \left(\frac{\overline{\alpha}^2}{\underline{\alpha}(M-1)}-\underline{\alpha}\right), \ \Gamma^{\gamma}_V = \overline{\gamma}\left(\frac{\overline{\alpha}}{\underline{\alpha}(M-1)}+1\right). \nonumber
\end{equation}

From the above inequality it follows that $\Gamma_V \leq -\tilde{\epsilon}$ and $\Gamma^{\gamma}_V \leq \overline{\gamma}\left((\underline{\alpha}-\tilde{\epsilon})/\overline{\alpha} + 1\right)$. Hence,
\begin{equation}
V^{t+1}_M - V^{t}_M \leq -\tilde{\epsilon} \sigma(x) + \overline{\gamma}\left(\frac{\underline{\alpha}-\tilde{\epsilon}}{\overline{\alpha}} + 1\right) \sum_{k=t}^{t^e} \norm{w_k}^2_2. \nonumber
\end{equation}

Then using the condition that $V^t_M \leq \overline{\alpha}\sigma(x_t) + \overline{\gamma} \sum_{k=t}^{t^e-1} \norm{w_t}^2_2$ we get
\begin{align}
V^{t+1}_M - V^{t}_M & \leq -\frac{\tilde{\epsilon}}{\overline{\alpha}}\left( V^t_M - \overline{\gamma} \sum_{k=t}^{t^e-1} \norm{w_k}^2_2\right) + \overline{\gamma}\left(\frac{\underline{\alpha}-\tilde{\epsilon}}{\overline{\alpha}} + 1\right) \sum_{k=t}^{t^e} \norm{w_k}^2_2 \nonumber\\
& = -\frac{\tilde{\epsilon}}{\overline{\alpha}} V^t_M + \overline{\gamma}\left(\frac{\underline{\alpha}}{\overline{\alpha}} + 1\right) \sum_{k=t}^{t^e} \norm{w_k}^2_2. \nonumber
\end{align}

Pick $\epsilon$ such that $0 < \epsilon \leq \tilde{\epsilon}/\overline{\alpha}$. Then
\begin{equation}
V^{t+1}_M - V^{t}_M \leq - \epsilon V^t_M + \overline{\gamma}\left(\frac{\underline{\alpha}}{\overline{\alpha}} + 1\right) \sum_{k=t}^{t^e} \norm{w_k}^2_2. \nonumber
\end{equation}

Let $a = (1- \epsilon)$. Note that $a < 1$. Let $b = \overline{\gamma}\left(\frac{\underline{\alpha}}{\overline{\alpha}} + 1\right)$. Then
\begin{equation}
V^{t+1}_M \leq  a V^t_M + b \sum_{k=t}^{t^e} \norm{w_k}^2_2. \nonumber
\end{equation}

Then, 
\begin{align}
& V^{t+2}_M \leq  a V^{t+1}_M + b \sum_{k=t+1}^{t^e+1} \norm{w_k}^2_2 \leq a\left(a V^t_M + b \sum_{k=t}^{t^e} \norm{w_k}^2_2\right) + b \sum_{k=t+1}^{t^e+1} \norm{w_k}^2_2 \nonumber\\
& = a^2V^{t}_M + ab \norm{w_t}^2_2 + b(1+a)\sum_{k=t+1}^{t^e} \norm{w_k}^2_2 + b \norm{w_{t^e+1}}^2_2\nonumber.
\end{align}

Similarly,
\begin{align}
V^{t+3}_M & \leq  a V^{t+2}_M + b \sum_{k=t+2}^{t^e+2} \norm{w_k}^2_2 \leq a^3V^{t}_M + a^2b \norm{w_t}^2_2 + ab(1+a)\norm{w_{t+1}}^2_2 \nonumber \\
& + b(1+a+a^2)\sum_{k=t+2}^{t^e} \norm{w_k}^2_2 + b(1+a) \norm{w_{t^e+1}}^2_2 + b\norm{w_{t^e+2}}^2_2. \nonumber
\end{align}

Extending the previous equation we get that for $H \leq M$
\begin{align}
& V^{t+H}_M \leq a^{H}V^{t}_M + b\sum_{k=0}^{H-2}\left(\left(\sum_{j=0}^{k} a^{H-1-j} \right)\norm{w_{t+k}}^2_2\right)+ b\left(\sum_{j = 0}^{H-1}a^j\right)\left(\sum_{k=H-1}^{M-1} \norm{w_{t+k}}^2_2\right)\nonumber \\
& + b\sum_{k = M}^{H+M-2} \left(\left(\sum_{j = 0}^{H+M-2-k}a^{j}\right) \norm{w_{t+k}}^2_2\right).
\label{eq:lem2-ind-init}
\end{align}

Next, we prove the upper bound for the case $H > M$ by induction. We hypothesize that $V^{t+H}_M$ for $H \geq M$ is given by
\begin{align}
& V^{t+H}_M \leq a^H V^{t}_M + b\sum_{k=0}^{M-1}\left(\left(\sum_{j=0}^{k} a^{H-1-j}\right)\norm{w_{t+k}}^2_2\right) \nonumber \\
& + b\frac{1-a^M}{1-a}\sum_{k=M}^{H-1}a^{H-k-1}\norm{w_{t+k}}^2_2 + b\sum_{k = H}^{H+M-2} \left(\left(\sum_{j = 0}^{H+M-2-k}a^j\right) \norm{w_{t+k}}^2_2\right) 
\label{eq:lem2-ind-hyp}
\end{align}

From Lemma \ref{lem:Lyapunov-existence} we know that
\begin{equation}
V^{t+H+1}_M \leq a V^{t+H}_M + b \sum_{k=t+H}^{t+H+M-1} \norm{w_k}^2_2 \nonumber 
\end{equation}

Substituting Eq. \eqref{eq:lem2-ind-hyp}
\begin{align}
& V^{t+H+1}_M \leq a \left(a^H V^{t}_M + b\sum_{k=0}^{M-1}\left(\left(\sum_{j=0}^{k} a^{H-1-j}\right)\norm{w_{t+k}}^2_2\right) \right. \nonumber \\
& \left. + b\frac{1-a^M}{1-a}\sum_{k=M}^{H-1}a^{H-k-1}\norm{w_{t+k}}^2_2 + b\sum_{k = H}^{H+M-2} \left(\left(\sum_{j = 0}^{H+M-2-k}a^j\right) \norm{w_{t+k}}^2_2\right) \right) + b \sum_{k=t+H}^{t+H+M-1} \norm{w_k}^2_2 \nonumber \nonumber \\
& = a^{H+1} V^{t}_M + b\sum_{k=0}^{M-1}\left(\left(\sum_{j=0}^{k} a^{H-j}\right)\norm{w_{t+k}}^2_2\right) + b\frac{1-a^M}{1-a}\sum_{k=M}^{H-1}a^{H-k}\norm{w_{t+k}}^2_2 \nonumber \\
& + b\sum_{k = H}^{H+M-2} \left(\left(\sum_{j = 0}^{H+M-2-k}a^{j+1}\right) \norm{w_{t+k}}^2_2\right) + b \sum_{k=t+H}^{t+H+M-1} \norm{w_k}^2_2 \nonumber \\
& = a^{H+1} V^{t}_M + b\sum_{k=0}^{M-1}\left(\left(\sum_{j=0}^{k} a^{H-j}\right)\norm{w_{t+k}}^2_2\right) + b\frac{1-a^M}{1-a}\sum_{k=M}^{H}a^{H-k}\norm{w_{t+k}}^2_2 \nonumber\\
& + b\sum_{k = H+1}^{H+M-1} \left(\left(\sum_{j = 0}^{H+M-1-k}a^{j}\right) \norm{w_{t+k}}^2_2\right). \nonumber
\end{align}

From Eq. \eqref{eq:lem2-ind-init} the hypothesis Eq. \eqref{eq:lem2-ind-hyp} is true for $H = M$. Hence, by induction Eq. \eqref{eq:lem2-ind-hyp} is true for all $H \geq M$. We can simplify Eq. \eqref{eq:lem2-ind-hyp} as
\begin{align}
& V^{t+H}_M \leq a^H V^{t}_M + ba^{H-M}\left(\sum_{k=0}^{M-1}\frac{a^{M-k-1}\left(1-a^{k+1}\right)}{1-a}\norm{w_{t+k}}^2_2\right) + b\frac{1-a^M}{1-a}\sum_{k=M}^{H-1}a^{H-k-1}\norm{w_{t+k}}^2_2 \nonumber \\ 
& + b\sum_{k = H}^{H+M-2} \left(\left(\sum_{j = 0}^{H+M-2-k}a^j\right) \norm{w_{t+k}}^2_2\right) \nonumber\\
& \leq a^H V^{t}_M + \frac{ba^{H-M}}{1-a}\left(\sum_{k=0}^{M-1}\norm{w_{t+k}}^2_2\right) + \frac{b}{1-a}\sum_{k=M}^{H-1}a^{H-k-1}\norm{w_{t+k}}^2_2 + \frac{b}{1-a}\sum_{k = H}^{H+M-2} \norm{w_{t+k}}^2_2. \nonumber 
\end{align}

Since $V^{t+H}_M \geq c_{t_H}(x_{t_H},\kappa_M(x_{t_H}))$. Then using the condition that $V^t_M \leq \overline{\alpha}\sigma(x_t) + \overline{\gamma} \sum_{k=t}^{t^e} \norm{w_{k}}^2_2$ we get
\begin{align}
& c_{t_H}(.,.) \leq a^H\overline{\alpha}\sigma(x_t) + \left(\frac{ba^{H-M}}{1-a}+a^H\overline{\gamma}\right)\left(\sum_{k=0}^{M-1}\norm{w_{t+k}}^2_2\right) \nonumber \\
& + \frac{b}{(1-a)}\sum_{k=M}^{H-1}a^{H-k-1}\norm{w_{t+k}}^2_2 + \frac{b}{(1-a)}\sum_{k = H}^{H+M-2} \norm{w_{t+k}}^2_2. \nonumber 
\end{align}

Thus, for any $0 < \epsilon \leq \tilde{\epsilon}/\overline{\alpha}$, $a = 1-\epsilon$
\begin{align}
& M_\lambda = a^H\overline{\alpha}; \ \lambda = -\ln{a}; M_{w,j} =  \left(\frac{ba^{H-M}}{1-a}+a^H\overline{\gamma}\right), t \leq j \leq t^e; \nonumber \\
& M_{w,j} = \frac{b}{(1-a)}a^{t+H-j-1}, t^e+1 \leq j \leq t+H-1; \nonumber \\
& M_{w,j} = \frac{b}{(1-a)}, t+H \leq j \leq t+H+M-2 \ \blacksquare \nonumber
\end{align}

\section*{Appendix C: Theorem \ref{thm:cost-regret}}

In the following we denote the cost incurred at time step $t$ succinctly by $c_t$ ignoring its arguments. Let $H \geq M$. Given that $M > \overline{\alpha}^2/\underline{\alpha}^2+1$, Lemma \ref{lem:cost-bound} is applicable to bound $c_{H+1}$. Hence, for any $a$ s.t. $1 > a \geq 1 - \tilde{\epsilon}/\overline{\alpha}, 0 < \tilde{\epsilon} < \underline{\alpha} - \frac{\overline{\alpha}^2}{(M-1)\underline{\alpha}}$, the stage cost of the closed loop system at $H+1$ is given by, 
\begin{align}
& c_{H+1} \leq a^H\overline{\alpha}\sigma(x_1) + \left(\frac{ba^{H-M}}{1-a}+a^H\overline{\gamma}\right)\left(\sum_{k=0}^{M-1}\norm{w_{1+k}}^2_2\right) \nonumber \\
& + \frac{b}{(1-a)}\sum_{k=M}^{H-1}a^{H-k-1}\norm{w_{1+k}}^2_2 + \frac{b}{1-a}\sum_{k = H}^{H+M-2} \norm{w_{t+k}}^2_2. \nonumber 
\end{align}

Hence, 
\begin{align}
& \sum_{H=M}^{T-1} c_{H+1} \leq \frac{\overline{\alpha}a^M}{1-a} \sigma(x_1) + \left(\frac{b}{(1-a)^2}+\frac{a^M\overline{\gamma}}{1-a}\right)\left(\sum_{k=0}^{M-1}\norm{w_{1+k}}^2_2\right)  \nonumber \\
& + \frac{b}{(1-a)}\sum_{H=M}^{T-1} \sum_{k=M}^{H-1}a^{H-k-1}\norm{w_{1+k}}^2_2 + \frac{b}{1-a}\sum_{H=M}^{T-1} \sum_{k = H}^{H+M-2} \norm{w_{t+k}}^2_2. \nonumber 
\end{align}

The first two terms are $O(1)$. Combining the terms corresponding to each disturbance realization $w_t$ we get
\begin{equation}
\sum_{H=M}^{T-1} c_{H+1} \leq O(1) + \frac{b\left((M-1)(1-a)+1\right)}{(1-a)^2}\sum_{H = M}^{T} \norm{w_H}^2_2 + \frac{b(M-1)}{1-a}\sum_{H=T}^{T+M-2} \norm{w_H}^2_2\nonumber 
\end{equation}

The last term is also $O(1)$. Hence,
\begin{equation}
\sum_{H=M}^{T-1} c_{H+1} \leq O(1) + \frac{b\left((M-1)(1-a)+1\right)}{(1-a)^2}\sum_{H = M}^{T} \norm{w_H}^2_2. \nonumber 
\end{equation}

From Eq. \eqref{eq:lem2-ind-init} in the proof of Lemma \ref{lem:cost-bound} we get that $\sum_{H=1}^{M-1} C_H \leq O(1)$. Then for $\gamma_c = \frac{b\left((M-1)(1-a)+1\right)}{(1-a)^2}$ we get
\begin{equation}
\sum_{H=1}^{T-1} c_{H} - \gamma_c\sum_{H = M}^{T} \norm{w_H}^2_2 \leq O(1) \ \blacksquare \nonumber 
\end{equation}

\section*{Appendix D: Disturbance is not Accessible}

In this setting $M_c = M, M_w = 0$. The receding horizon online controller for this case optimizes the worst-case cost-to-go instead of the cost-to-go for the realized disturbance:
\begin{align}
& \underset{\tilde{u}_{t:t+M-1}}{\inf} \underset{\tilde{w}_{t:t+M-1}}{\sup} \sum_{j = t}^{t+M-1} c_j(\tilde{x}_j, \tilde{u}_j) \ \text{s.t.} \ \tilde{x}_{j+1} = f(\tilde{x}_j, \tilde{u}_j, \tilde{w}_j), , \nonumber \\
& \ \tilde{x}_t = x_t, \ \tilde{u}_j \in \mathcal{U}, \ \tilde{w}_j \in \mathcal{W}.
\label{eq:cont-opt-no-distaccess}
\end{align}
Given the fact that the constraint $\mathcal{W}$ is compact, the solution to the inner optimization exists and is continuous by an argument similar to Eq. \eqref{eq:cont-opt-distaccess} (see Proof of Lemma \ref{lem:Lipschitz}). Applying the same argument again for the outer optimization we find that the solution to the outer optimization exists and is continuous. 

While RHC algorithms are complex because of the optimization, min-max optimization further exacerbate the difficulty because of the bi-level nature of the optimization. Min-max optimization of general non-convex and non-concave objective functions are known to be hard problems \cite{daskalakis2020complexity}. It was shown in \cite{daskalakis2020complexity} that finding an approximate local min-max point of large enough approximation is guaranteed to exist, but finding one such point is PPAD-complete. When the objective function $L(x,y)$ is a convex-concave function, i.e., L is convex in x for
all y and it is concave in y for all x, the problem $\min_x\max_y L(x,y)$ with constraints is guaranteed to have a solution, under compactness of the constraint set \cite{rosen1965existence}, while computing a solution is amenable to convex programming. In fact, if L is $L-$smooth, the problem can be solved via first-order methods, and achieve an approximation error of $\text{poly}(L, 1/T)$ in $T$ iterations; see e.g. \cite{nemirovski2004interior}. When the function is strongly convex and strongly concave, the rate becomes geometric \cite{facchinei2007finite}.

We denote the solution computed by the optimization Eq. \eqref{eq:cont-opt-no-distaccess} as $\text{MPC}^t_\mathcal{W}$ and $W^{*}_{t:t+M-1}$. The control input $u_t$ is set as the first element of the sequence $\text{MPC}^t_\mathcal{W}$. We succinctly denote this by $\kappa_{\mathcal{W},M}(x_t)$. The complete algorithm for this case is given in Algorithm \ref{alg:online-control-no-distaccess}. 

\begin{algorithm}[]
\begin{algorithmic}[1]
\STATE \textbf{Input}: $M$
\FOR{t = 1,...,T}
\STATE Observe $y_t$, $\{c_j(.,.)\}_{t \leq j \leq t+M-1}$ 
\STATE Solve the optimization Eq. \eqref{eq:cont-opt-no-distaccess} to obtain $\tilde{u}_{t:t+M-1}$
\STATE Apply $\tilde{u}_t$ to the system
\ENDFOR
\end{algorithmic}
\caption{Online Learning Robust Control (Known System and no Preview of Disturbance)}
\label{alg:online-control-no-distaccess}
\end{algorithm}

Let 
\begin{align}
& V^t_{\mathcal{W},M}(x_t) = J^t_M(x_t, \tilde{u}_{t:t+M-1}, \tilde{w}_{t:t+M-1}), \nonumber \\
& J^t_M(x_t, \tilde{u}_{t:t+M-1}, \tilde{w}_{t:t+M-1}) =  \sum_{j = t}^{t+M-1} c_j(\tilde{x}_j, \tilde{u}_j) \ \text{s.t.} \ \tilde{x}_{j+1} = f(\tilde{x}_j, \tilde{u}_j, \tilde{w}_j), \tilde{x}_t = x_t. \nonumber
\end{align}

By the existence of the solution $\text{MPC}^t_\mathcal{W}$ this function is well defined. In the next lemma, we characterize the incremental change of the value of this function as the closed loop system evolves from $t$ to $t+1$. As in the previous case, we establish the result for the class of systems that satisfy $V^t_{\mathcal{W},M}(x_t) \leq \overline{\alpha}_\mathcal{W}\sigma(x_t) + \overline{\gamma}_\mathcal{W} \max_{w\in\mathcal{W}} \norm{w}^2_2$. In contrast to the previous case, the bound of this function is specified in terms of the maximum disturbance. Given the min-max nature of the optimization where the maximization is carried out overall all disturbance realizations this bound naturally follows from the bound assumed on $V^t_M(.)$ in the previous case. It is easy to see that $\overline{\gamma}$ and $\overline{\gamma}_\mathcal{W}$ are related by $\overline{\gamma}_\mathcal{W} = \overline{\gamma}M$. We once again note that this class of systems trivially includes the strongly convex quadratic cost functions for linear dynamical systems.
\begin{lemma}
Suppose $t^e = t+M-1$. Then there exists constants $\underline{\alpha}_{\mathcal{W},V}, \Gamma^{\gamma}_{\mathcal{W},V} > 0$ and $\Gamma_{\mathcal{W},V}$ such that for each $M \geq 2$ and $x_t \in \mathbb{R}^n$, for the closed-loop system $x_{t+1} = f(x_t, \kappa^T_{\mathcal{W},M}(x_t), w_t)$, $V^t_{\mathcal{W},M}(x_t) \geq \underline{\alpha}_{\mathcal{W},V}\sigma(x_t)$ and
\[
V^{t+1}_{\mathcal{W},M}(x_{t+1}) - V^t_{\mathcal{W},M}(x_t) \leq \Gamma_{\mathcal{W},V} \sigma(x_t) + \Gamma^{\gamma}_{\mathcal{W},V} \max_{w\in\mathcal{W}} \norm{w}^2_2,\]
where $\underline{\alpha}_{\mathcal{W},V} = \underline{\alpha}$, $\Gamma_{\mathcal{W},V} = \left(\frac{\overline{\alpha}^2_\mathcal{W}}{(M-1)\underline{\alpha}} -\underline{\alpha}\right), \ \Gamma^{\gamma}_{\mathcal{W},V} = \overline{\gamma}_\mathcal{W}\left(\frac{\overline{\alpha}_\mathcal{W}}{(M-1)\underline{\alpha}}+1\right)$.
\label{lem:Lyapunov-existence-nodist}
\end{lemma}

{\em Proof}:
We note that
\begin{equation}
V^t_{\mathcal{W},M}(x_{t}) \geq c_t(x_t, \kappa_{\mathcal{W},M}(x_t)) \geq \underline{\alpha}\sigma(x_t). \nonumber
\end{equation}

The last inequality follows directly from Assumption \ref{ass:stagecost}. Hence, $\underline{\alpha}_{\mathcal{W},V} = \underline{\alpha}$. We introduce some notation for simplifying the presentation of the analysis. If $\text{MPC}^t_\mathcal{W}= \{\tilde{u}^*_t, \tilde{u}^*_{t+1},...,\tilde{u}^*_{t^e}\}$, then let $\text{MPC}^t_\mathcal{W}[i:j] = \{\tilde{u}^*_{t+i-1},..., \tilde{u}^*_{t+j-1}\}, \ i \leq j$. We denote $\text{MPC}^t_\mathcal{W}$ by $\text{MPC}^t_\mathcal{W}$, $W_{t:t^e}$ by $W_{t}$, and $W^{*}_{t:t^e}$ by $W^{*}_{t}$. Similarly, let $W_t[i:j] = W_{t+i-1:t+j-1}, \ W^{*}_t[i:j] = W^{*}_{t+i-1:t+j-1}$

Let $\phi^t(k, x_t, \text{M}^t_\mathcal{W}, W_t)$ denote the state the system evolves to at time $t+k$ starting at $x_t$, under the control sequence $\text{M}^t_{\mathcal{W}}[1:k]$ and disturbance $W_{t}$. From the definition of $V^{t}_{\mathcal{W},M}(x_{t})$ it follows that
\begin{align}
& V^{t+1}_{\mathcal{W},M}(x_{t+1}) - V^{t}_{\mathcal{W},M}(x_{t}) = J^{t+1}_M(x_{t+1}, \text{M}^{t+1}_\mathcal{W}, W^{*}_{t+1}) - J^{t}_M(x_t, \text{M}^t_\mathcal{W}, W^{*}_{t}) \nonumber \\
& = \sum_{k = 0}^{M-1} c_{t+1+k}(\phi^{t+1}(k, x_{t+1}, \text{M}^{t+1}_\mathcal{W}, W^{*}_{t+1}), \text{M}^{t+1}_\mathcal{W}[k+1])\nonumber \\
&- \sum_{k = 0}^{M-1} c_{t+k}(\phi^t(k, x_t, \text{M}^{t}_\mathcal{W}, W^*_t), \text{M}^{t}_\mathcal{W}[k+1]). \nonumber
\end{align}

Define 
\begin{align}
& \tilde{\text{M}}^{t+1}_\mathcal{W} = [\text{M}^{t}_\mathcal{W}[2:j+1], \ \tilde{U}^{M-j}_j], \ \tilde{U}^{M-j}_j = \{\tilde{u}_1,...,\tilde{u}_{M-j}\}, \nonumber \\
& \tilde{W}_{t+1} = [W^{*}_{t+1}[1:j], \ \tilde{W}^{M-j}_j], \ \tilde{W}^{M-j}_j = \{\tilde{w}_1,...,\tilde{w}_{M-j}\}. \nonumber 
\end{align}

Let
\begin{align}
& \{\tilde{U}^{*,M-j}_j, \tilde{W}^{*,M-j}_j\} = \underset{\tilde{U}^{M-j}_j}{\text{arginf}} \ \underset{ \tilde{W}^{M-j}_j}{\text{argsup}} \sum_{k=0}^{j-1} c_{t+1+k}(\phi^{t+1}(k, x_{t+1}, \tilde{\text{M}}^{t+1}_\mathcal{W}, \tilde{W}), \tilde{\text{M}}^{t+1}_\mathcal{W}[k+1])  \nonumber \\
& + J^{t+1+j}_{M-j}(\phi^{t+1}(j, x_{t+1}, \tilde{\text{M}}^{t+1}_\mathcal{W}, \tilde{W}), \tilde{U}^{M-j}_j ,\tilde{W}^{M-j}_j). \nonumber 
\end{align}

Let $\tilde{W}^{*,j}_{t+1} = [W^{*}_{t+1}[1:j], \ \tilde{W}^{*,M-j}_j]$. Given that $[w_t, \tilde{W}^{*,j}_{t+1}[1:M-1]]$ is not the worst-case disturbance realization for the system starting at $x_t$ and for the control sequence $\text{M}^{t}_\mathcal{W}$. Hence,
\begin{align}
& V^{t+1}_{\mathcal{W},M}(x_{t+1}) - V^{t}_{\mathcal{W},M}(x_{t}) \leq \sum_{k = 0}^{M-1} c_{t+1+k}(\phi^{t+1}(k, x_{t+1}, \text{M}^{t+1}_\mathcal{W}, W^{*}_{t+1}), \text{M}^{t+1}_\mathcal{W}[k+1]) \nonumber \\
& - c_t(x_t,\text{M}^{t}_\mathcal{W}[1]) - \sum_{k = 0}^{M-2} c_{t+1+k}(\phi^{t+1}(k, x_{t+1}, \text{M}^{t}_\mathcal{W}[2:k+1],  \tilde{W}^{*,j}_{t+1}[1:M-1]), \text{M}^{t}_\mathcal{W}[k+2]). \nonumber
\end{align}

Let $\tilde{\text{M}}^{*,j}_\mathcal{W} =  [\tilde{\text{M}}^{t+1}_\mathcal{W}[1:j],\tilde{U}^{*,M-j}_j]$. From the definitions of $\text{M}^{t+1}_\mathcal{W}, \ W^{*}_{t+1}, \ \tilde{\text{M}}^{*,j}_\mathcal{W}, \ \tilde{W}^{*,j}_{t+1}$ it follows that
\begin{align}
& \sum_{k = 0}^{M-1} c_{t+1+k}(\phi^{t+1}(k, x_{t+1}, \text{M}^{t+1}_\mathcal{W}, W^{*}_{t+1}), \text{M}^{t+1}_\mathcal{W}[k+1]) \nonumber \\
& \leq \sum_{k = 0}^{M-1} c_{t+1+k}(\phi^{t+1}(k, x_{t+1}, \tilde{\text{M}}^{*,j}_\mathcal{W}, W^{*}_{t+1}), \tilde{\text{M}}^{*,j}_\mathcal{W}[k+1]) \nonumber \\
& \leq \sum_{k = 0}^{M-1} c_{t+1+k}(\phi^{t+1}(k, x_{t+1}, \tilde{\text{M}}^{*,j}_\mathcal{W}, \tilde{W}^{*,j}_{t+1}), \tilde{\text{M}}^{*,j}_\mathcal{W}[k+1]). \nonumber
\end{align}

Using the above inequality we further get
\begin{align}
& V^{t+1}_{\mathcal{W},M}(x_{t+1}) - V^{t}_{\mathcal{W},M}(x_{t}) \leq \sum_{k = 0}^{M-1} c_{t+1+k}(\phi^{t+1}(k, x_{t+1}, \tilde{\text{M}}^{*,j}_\mathcal{W}, \tilde{W}^{*,j}_{t+1}), \tilde{\text{M}}^{*,j}_\mathcal{W}[k+1]) \nonumber \\
& - c_t(x_t,\text{M}^{t}_\mathcal{W}[1]) - \sum_{k = 0}^{M-2} c_{t+1+k}(\phi^{t+1}(k, x_{t+1}, \text{M}^{t}_\mathcal{W}[2:k+1],  \tilde{W}^{*,j}_{t+1}[1:M-1]), \text{M}^{t}_\mathcal{W}[k+2]). \nonumber
\end{align}

Using the definitions of $\tilde{\text{M}}^{*,j}_\mathcal{W}$ and $\tilde{W}^{*,j}_{t+1}$, we get
\begin{align}
& V^{t+1}_{\mathcal{W},M}(x_{t+1}) - V^{t}_{\mathcal{W},M}(x_{t}) \leq \sum_{k=0}^{j-1} c_{t+1+k}(\phi^{t+1}(k, x_{t+1}, \tilde{\text{M}}^{t+1}_\mathcal{W}, \tilde{W}^{*,j}_{t+1}), \tilde{\text{M}}^{t+1}_\mathcal{W}[k+1]) \nonumber \\
& + J^{t+1+j}_{M-j}(\phi^{t+1}(j, x_{t+1}, \tilde{\text{M}}^{t+1}_\mathcal{W}, \tilde{W}^{*,j}_{t+1}), \tilde{U}^{*,M-j}_j ,\tilde{W}^{*,M-j}_j) - c_t(x_t,\text{M}^{t}_\mathcal{W}[1]) \nonumber \\
& - \sum_{k = 0}^{M-2} c_{t+1+k}(\phi^{t+1}(k, x_{t+1}, \text{M}^{t}_\mathcal{W}[2:k+1],  \tilde{W}^{*,j}_{t+1}[1:M-1]), \text{M}^{t}_\mathcal{W}[k+2]). \nonumber 
\end{align}

Then for any $j \in \{0,...,M-1\}$, we get
\begin{align}
&  V^{t+1}_{\mathcal{W},M}(x_{t+1}) - V^{t}_{\mathcal{W},M}(x_{t}) \leq J^{t+1+j}_{M-j}(\phi^{t+1}(j, x_{t+1}, \tilde{\text{M}}^{t+1}_\mathcal{W}, \tilde{W}^{*,j}_{t+1}), \tilde{U}^{*,M-j}_j ,\tilde{W}^{*,M-j}_j) \nonumber \\
& - c_t(x_t,\text{M}^{t}_\mathcal{W}[1]). \nonumber 
\end{align}

By definition 
\begin{equation}
J^{t+1+j}_{M-j}(\phi^{t+1}(j, x_{t+1}, \tilde{\text{M}}^{t+1}_\mathcal{W}, \tilde{W}^{*,j}_{t+1}), \tilde{U}^{*,M-j}_j ,\tilde{W}^{*,M-j}_j) = V^{t+1+j}_{\mathcal{W},M-j}(\phi^{t+1}(j, x_{t+1}, \tilde{\text{M}}^{t+1}_\mathcal{W}, \tilde{W}^{*,j}_{t+1})). \nonumber 
\end{equation}

Hence for any $j \in \{0,...,M-1\}$, 
\begin{equation}
V^{t+1}_{\mathcal{W},M}(x_{t+1}) - V^{t}_{\mathcal{W},M}(x_{t}) \leq V^{t+1+j}_{\mathcal{W},M-j}(\phi^{t+1}(j, x_{t+1}, \tilde{\text{M}}^{t+1}_\mathcal{W}, \tilde{W}^{*,j}_{t+1})) - c_t(x_t,\text{M}^{t}_\mathcal{W}[1]). \nonumber   
\end{equation}

By definition
\begin{equation}
V^{t+1+j}_{\mathcal{W},M-j}(.) \leq V^{t+1+j}_{\mathcal{W},M}(.) \leq  \overline{\alpha}_\mathcal{W}\sigma(.) + \overline{\gamma}_\mathcal{W} \max_{w\in\mathcal{W}} \norm{w}^2_2. \nonumber   
\end{equation}

Using the fact that $\tilde{\text{M}}^{*,j}_\mathcal{W} = [\text{M}^{t}_\mathcal{W}[2:j+1],\tilde{U}^{*,M-j}_j]$ and $\tilde{W}^{*,j}_{t+1} = [W^{*}_{t+1}[1:j], \ \tilde{W}^{*,M-j}_j]$, it follows that 
\beq 
\phi^{t+1}(j, x_{t+1}, \tilde{\text{M}}^{t+1}_\mathcal{W}, \tilde{W}^{*,j}_{t+1}) = \phi^{t+1}(j, x_{t+1}, \text{M}^{t}_\mathcal{W}[2:j+1], W^{*}_{t+1}[1:j]). \nonumber 
\eeq 

Hence,
\begin{align}
& V^{t+1}_{\mathcal{W},M}(x_{t+1}) - V^{t}_{\mathcal{W},M}(x_{t}) \leq \overline{\alpha}_\mathcal{W}\sigma(\phi^{t+1}(j, x_{t+1}, \text{M}^{t}_\mathcal{W}[2:j+1], W^{*}_{t+1}[1:j])) \nonumber \\ 
& + \overline{\gamma}_\mathcal{W} \max_{w\in\mathcal{W}} \norm{w}^2_2- c_t(x_t,\text{M}^{t}_\mathcal{W}[1]). \nonumber 
\end{align}

By Assumption \ref{ass:stagecost} (similar to the step in the Lemma \ref{lem:Lyapunov-existence}) we have that
\begin{align}
& \sum_{k=0}^{M-2} \underline{\alpha} \sigma(\phi^{t+1}(k, x_{t+1}, \text{M}^{t}_\mathcal{W}[2:k+1], W^{*}_{t+1}[1:k])) \leq c_t(x_t,\text{M}^{t}_\mathcal{W}[1]) \nonumber \\
& + \sum_{k=0}^{M-2} c_{t+1+k}(\phi^{t+1}(k, x_{t+1}, \text{M}^{t}_\mathcal{W}[2:M], W^{*}_{t+1}), \text{M}^{t}_\mathcal{W}[k+2]). \nonumber
\end{align}

By definition $W^{*}_t$ is the worst-case disturbance realization for the control sequence $\text{M}^{t}_\mathcal{W}$ and the system starting $x_t$. Hence,
\begin{align}
& c_t(x_t,\text{M}^{t}_\mathcal{W}[1]) + \sum_{k=0}^{M-2} c_{t+1+k}(\phi^{t+1}(k, x_{t+1}, \text{M}^{t}_\mathcal{W}[2:M], W^{*}_{t+1}), \text{M}^{t}_\mathcal{W}[k+2]) \nonumber \\
& \leq \sum_{k = 0}^{M-1} c_{t+k}(\phi^{t}(k, x_{t}, \text{M}^{t}_\mathcal{W}, W^{*}_{t}), \text{M}^{t}_\mathcal{W}[k+1]) = V^t_{\mathcal{W},M} \leq \overline{\alpha}_\mathcal{W} \sigma(x_t) + \overline{\gamma}_\mathcal{W}\max_{w\in\mathcal{W}}\norm{w}^2_2, \nonumber
\end{align}
Where the last inequality follows from $V^t_{\mathcal{W},M}(.) \leq \overline{\alpha}_\mathcal{W} \sigma(.) + \overline{\gamma}_\mathcal{W}\max_{w\in\mathcal{W}}\norm{w}^2_2$. Using this inequality we get that
\begin{equation}
\sum_{k=0}^{M-2} \underline{\alpha} \sigma(\phi^{t+1}(k, x_{t+1}, \text{M}^{t}_\mathcal{W}[2:k+1], W^{*}_{t+1}[1:k])) \leq \overline{\alpha}_\mathcal{W} \sigma(x_t) + \overline{\gamma}_\mathcal{W}\max_{w\in\mathcal{W}}\norm{w}^2_2. \nonumber 
\end{equation}

Hence there exists a $j^{*} \in \{0,...,M-2\}$ such that
\begin{align}
& \sigma(\phi^{t+1}(j^{*}, x_{t+1}, \text{M}^{t}_\mathcal{W}[2:j^{*}+1], W^{*}_{t+1}[1:j^{*}])) \nonumber \\
& \leq \frac{1}{(M-1)\underline{\alpha}}\left(\overline{\alpha}_\mathcal{W}\sigma(x_t) + \overline{\gamma}_\mathcal{W}\max_{w\in\mathcal{W}}\norm{w}^2_2\right). \nonumber   
\end{align}

Hence, 
\begin{align}
& V^{t+1}_{\mathcal{W},M}(x_{t+1}) - V^{t}_{\mathcal{W},M}(x_{t}) \leq \left(\frac{\overline{\alpha}^2_\mathcal{W}}{(M-1)\underline{\alpha}} - \underline{\alpha}\right)\sigma(x_t) \nonumber \\
& + \overline{\gamma}_\mathcal{W}\left(\frac{\overline{\alpha}_\mathcal{W}}{(M-1)\underline{\alpha}}+1\right) \max_{w\in\mathcal{W}} \norm{w}^2_2 \ \blacksquare\nonumber   
\end{align}

Similar to the previous case two terms bound the incremental change in the value of function $V^t_{\mathcal{W},M}$: the first term is the contribution from the cost for the state $x_t$ and the second term is the contribution from the maximum possible energy for the disturbance recognizing that the factor $\overline{\gamma}_\mathcal{W}$ in $\Gamma^\gamma_{\mathcal{W},V}$ is $\leq \overline{\gamma}M$. Next we use the above lemma to characterize the performance of the online controller.
\begin{theorem}
Suppose the conditions of Lemma \ref{lem:Lyapunov-existence-nodist}. Then, for $T \geq M+1$, $M > \overline{\alpha}^2_\mathcal{W}/\underline{\alpha}^2 + 1$, $b = \overline{\gamma}\left(\frac{\underline{\alpha}}{\overline{\alpha}_\mathcal{W}} + 1\right)$, and any $a$ s.t. $1 > a \geq 1 - \tilde{\epsilon}/\overline{\alpha}_\mathcal{W}, 0 < \tilde{\epsilon} < \underline{\alpha} - \frac{\overline{\alpha}_\mathcal{W}^2}{(M-1)\underline{\alpha}}$, the total cost incurred by the closed-loop system $x_{t+1} = f(x_t, \kappa^T_{\mathcal{W},M}(x_t), w_t)$ satisfies
\begin{equation}
\sum_{t=1}^{T} c_{t}(x_{t},\kappa_{\mathcal{W},M}(x_{t})) - \gamma T \max_{w\in\mathcal{W}} \norm{w}^2_2  \leq O(1), \ \gamma \geq \gamma_{c,\mathcal{W}} \nonumber
\end{equation}
where $\gamma_{c,\mathcal{W}} = \frac{\overline{\gamma}_\mathcal{W}}{1-a}\left(\frac{\underline{\alpha}}{\overline{\alpha}_\mathcal{W}}+1\right)$.
\label{thm:cost-regret-nodist}
\end{theorem}

{\em Proof}: Since $M > \overline{\alpha}^2_\mathcal{W}/\underline{\alpha}^2+1$, there exists $\tilde{\epsilon} \in (0, \underline{\alpha})$ such that  such that 
\begin{equation}
M > \overline{\alpha}^2_\mathcal{W}/\left(\underline{\alpha}(\underline{\alpha}-\tilde{\epsilon})\right) + 1 \geq \overline{\alpha}_\mathcal{W}/\left(\underline{\alpha}-\tilde{\epsilon}\right)+1. \nonumber
\end{equation}

For convenience of presentation in the following we denote $V^{t}_{\mathcal{W},M}(x_{t})$ by $V^{t}_{\mathcal{W},M}$. From Lemma \ref{lem:Lyapunov-existence} it follows that
\begin{equation}
V^{t+1}_{\mathcal{W},M} - V^{t}_{\mathcal{W},M} \leq \left(\frac{\overline{\alpha}_\mathcal{W}^2}{\underline{\alpha}(M-1)}-\underline{\alpha}\right) \sigma(x) + \overline{\gamma}_\mathcal{W}\left(\frac{\overline{\alpha}_\mathcal{W}}{\underline{\alpha}(M-1)}+1\right) \max_{w\in\mathcal{W}} \norm{w}^2_2. \nonumber
\end{equation}

Hence,
\begin{equation}
V^{t+1}_{\mathcal{W},M} - V^{t}_{\mathcal{W},M} \leq -\tilde{\epsilon}\sigma(x) + \overline{\gamma}_\mathcal{W}\left(\frac{\overline{\alpha}_\mathcal{W}}{\underline{\alpha}(M-1)}+1\right) \max_{w\in\mathcal{W}} \norm{w}^2_2. \nonumber
\end{equation}

Then using the fact that $V^{t}_{\mathcal{W},M}(.) \leq \overline{\alpha}_\mathcal{W}\sigma(.) + \overline{\gamma}_\mathcal{W}\max_{w\in\mathcal{W}} \norm{w}^2_2$ we get
\begin{equation}
V^{t+1}_{\mathcal{W},M} - V^{t}_{\mathcal{W},M} \leq -\frac{\tilde{\epsilon}}{\overline{\alpha}_\mathcal{W}}\left(V^{t}_{\mathcal{W},M}(x_t) - \overline{\gamma}_\mathcal{W}\max_{w\in\mathcal{W}} \norm{w}^2_2\right) + \overline{\gamma}_\mathcal{W}\left(\frac{\overline{\alpha}_\mathcal{W}}{\underline{\alpha}(M-1)}+1\right) \max_{w\in\mathcal{W}} \norm{w}^2_2. \nonumber
\end{equation}

Then using the fact that $\underline{\alpha}-\tilde{\epsilon} > \overline{\alpha}^2_\mathcal{W}/\left(\underline{\alpha}(M-1)\right)$ we get
\begin{equation}
V^{t+1}_{\mathcal{W},M} - V^{t}_{\mathcal{W},M} \leq -\frac{\tilde{\epsilon}}{\overline{\alpha}_\mathcal{W}}V^{t}_{\mathcal{W},M}(x_t) + \overline{\gamma}_\mathcal{W}\left(\frac{\underline{\alpha}}{\overline{\alpha}_\mathcal{W}}+1\right) \max_{w\in\mathcal{W}} \norm{w}^2_2. \nonumber
\end{equation}

Pick $\epsilon$ s.t. $0 < \epsilon \leq \frac{\tilde{\epsilon}}{\overline{\alpha}_\mathcal{W}}$. Then
\begin{equation}
V^{t+1}_{\mathcal{W},M} - V^{t}_{\mathcal{W},M} \leq -\epsilon V^{t}_{\mathcal{W},M}(x_t) + \overline{\gamma}_\mathcal{W}\left(\frac{\underline{\alpha}}{\overline{\alpha}_\mathcal{W}}+1\right) \max_{w\in\mathcal{W}} \norm{w}^2_2. \nonumber
\end{equation}

Let $a = 1-\epsilon$. Then
\begin{equation}
V^{t+1}_{\mathcal{W},M} \leq a V^{t}_{\mathcal{W},M}(x_t) + \overline{\gamma}_\mathcal{W}\left(\frac{\underline{\alpha}}{\overline{\alpha}_\mathcal{W}}+1\right) \max_{w\in\mathcal{W}} \norm{w}^2_2. \nonumber
\end{equation}

For convenience we denote the cost incurred at time step $t+H$ just by $c_{t+H}$ ignoring its arguments. Then
\begin{align}
& V^{t+H}_{\mathcal{W},M} \leq a^H V^{t}_{\mathcal{W},M}(x_t) + \frac{\overline{\gamma}_\mathcal{W}(1-a^H)}{1-a}\left(\frac{\underline{\alpha}}{\overline{\alpha}_\mathcal{W}}+1\right) \max_{w\in\mathcal{W}} \norm{w}^2_2, \nonumber\\
& \text{i.e.,} \ c_{t+H} \leq a^H V^{t}_{\mathcal{W},M}(x_t) + \frac{\overline{\gamma}_\mathcal{W}(1-a^H)}{1-a}\left(\frac{\underline{\alpha}}{\overline{\alpha}_\mathcal{W}}+1\right) \max_{w\in\mathcal{W}} \norm{w}^2_2. \nonumber 
\end{align}

Then using the fact that $V^{t}_{\mathcal{W},M}(.) \leq \overline{\alpha}_\mathcal{W}\sigma(x) + \overline{\gamma}_\mathcal{W} \max_{w\in\mathcal{W}}\norm{w}^2_2$ we get
\begin{equation}
c_{t+H} \leq a^H\overline{\alpha}_\mathcal{W}\sigma(x_t) + \frac{ \overline{\gamma}_\mathcal{W}(1-a^{H+1})}{1-a}\left(\frac{\underline{\alpha}}{\overline{\alpha}_\mathcal{W}}+1\right) \max_{w\in\mathcal{W}} \norm{w}^2_2. \nonumber 
\end{equation}

Thus summing over all $H \in \{1,...,T\}$ we get
\begin{equation}
\sum_{H = 0}^{T-1} c_{1+H} - \frac{\overline{\gamma}_\mathcal{W}}{1-a}\left(\frac{\underline{\alpha}}{\overline{\alpha}_\mathcal{W}}+1\right)T\max_{w\in\mathcal{W}} \norm{w}^2_2 \leq \frac{\overline{\alpha}_\mathcal{W}}{1-a}\sigma(x_1) = O(1) \ \blacksquare \nonumber 
\end{equation}

We note two differences in the result: (i) the attenuation depends on $\overline{\gamma}_\mathcal{W}$ and (ii) the attenuation guarantee is given with respect to the maximum possible energy from the disturbances for the duration $T$, i.e., $T\max_{w\in\mathcal{W}} \norm{w}^2_2$. The controller's performance can only characterized in terms of the maximum possible energy of the disturbance since the controller is always regulating for the worst-case, naturally. Given that $\overline{\gamma}_\mathcal{W} = \overline{\gamma}M$, we can choose $M$ and $a$ such that, $2\tilde{\gamma}\left(1+\frac{\underline{\alpha}}{\overline{\alpha}}\right)\overline{\gamma} \gtrapprox \gamma_{c,\mathcal{W}} \geq 2\overline{\gamma}$. Thus the online RHC approach achieves an attenuation of $\mathcal{O}(\overline{\gamma})$ with respect to the maximum possible energy of the disturbance.

\section*{Appendix E: Proof of Lemma \ref{lem:Lipschitz}}

Here we prove that under Assumption \ref{ass:sys} the conditions required for Proposition 2.4, \cite{quincampoix2008parameterized} to hold are satisfied. Since $f, c_j$ are continuous, $\mathcal{W}$ is bounded, the objective function for the optimization in Eq. \eqref{eq:cont-opt-distaccess} is uniformly level bounded (see \cite{rockafellar2009variational} for definition) and by definition continuous functions are proper. This satisfies the assumption in Theorem 1.17, \cite{rockafellar2009variational} that the objective function is proper, lower semi-continuous and uniformly level bounded. Moreover by continuity of the functions there should exist a finite solution $\text{MPC}^t_{M}$ for all $\theta \in \Theta$, $x_t \in \mathbb{R}^n$ and all $W_{t:t^e}$ s.t. $w_t \in \mathcal{W}$. Hence the domain of the solution $\text{MPC}^t_{M}$ is $\{\theta,x,W_{t:t^e}: \theta \in \Theta, x\in\mathbb{R}^n, w_t\in \mathcal{W}\}$. Then by continuity of $f, c_j$ and point 3 of Theorem 1.17, \cite{rockafellar2009variational} the solution $\text{MPC}^t_{M}$ is continuous in $x_t, \theta$ and $W_{t:t^e}$. Hence Assumption A1 of proposition of Proposition 2.4, \cite{quincampoix2008parameterized} is satisfied. Since $c_t$ is strongly convex in the second argument it follows that Assumption A2 of Proposition 2.4, \cite{quincampoix2008parameterized} trivially holds. Finally, by (vii) of Assumption \ref{ass:sys} it follows that the Assumption A3 of Proposition 2.4, \cite{quincampoix2008parameterized} is satisfied. Thus, all the assumptions of Proposition 2.4, \cite{quincampoix2008parameterized} are satisfied. Hence, applying Proposition 2.4 we get the final result.

\section*{Appendix F: Proof of Lemma \ref{lem:Lyapunov-existence-unknownsystem}}

We note that
\begin{equation}
V^t_M(x_t,W_{t:t^e};\hat{\theta}) \geq c_t(x_t, \kappa_M(x_t;\hat{\theta})) \geq \underline{\alpha}\sigma(x_t). \nonumber
\end{equation}
The last inequality follows directly from Assumption \ref{ass:stagecost}. Hence, $\underline{\alpha}_V = \underline{\alpha} > 0$. We introduce some notation for simplifying the presentation of the analysis. If $\text{MPC}^t(W_{t:t^e};.) = \{\tilde{u}^*_t, \tilde{u}^*_{t+1},...,\tilde{u}^*_{t^e}\}$, then let $\text{MPC}^t( W_{t:t^e};.)[i:j] = \{\tilde{u}^*_{t+i-1},..., \tilde{u}^*_{t+j-1}\}, \ i \leq j$. We denote $\text{MPC}^t(W_{t:t^e};\theta)$ by $\text{M}^t_p$, and $W_{t:t^e}$ by $W_{t}$. 

Let $\phi^t(k, x_t, \text{M}^t_p, W_t)$ denote the state the system evolves to at time $t+k$ starting at $x_t$, under the control sequence $\text{M}^t_p[1:k]$ and disturbance $W_{t}$. From the definition of $V^{t}_M(x_{t},W_t;\hat{\theta})$ and Lemma \ref{lem:Lipschitz} it follows that
\begin{equation}
V^{t+1}_M(x_{t+1},W_{t+1};\hat{\theta}) - V^{t}_M(x_{t},W_{t};\hat{\theta}) \leq V^{t+1}_M(x_{t+1},W_{t+1};\theta) - V^{t}_M(x_{t},W_{t};\theta) + 2\alpha_V\norm{\hat{\theta}-\theta}_2. \nonumber  
\end{equation}

We now analyze the term $V^{t+1}_M(x_{t+1},W_{t+1};\theta) - V^{t}_M(x_{t},W_{t};\theta)$. By definition
\begin{align}
& V^{t+1}_M(x_{t+1},W_{t+1};\theta) - V^{t}_M(x_{t},W_{t};\theta) = J^{t+1}_M(x_{t+1}, \text{M}^{t+1}_p, W_{t+1}) - J^{t}_M(x_t, \text{M}^t_p, W_{t}) \nonumber \\
& = \sum_{k = 0}^{M-1} c_{t+1+k}(\phi^{t+1}(k, x_{t+1}, \text{M}^{t+1}_p, W_{t+1}), \text{M}^{t+1}_p[k+1]) \nonumber \\
& - \sum_{k = 0}^{M-1} c_{t+k}(\phi^t(k, x_t, \text{M}^t_p, W_t), \text{M}^t_p[k+1]). \nonumber
\end{align}

The second step just follows from the definition of $J^{t}_M(.,.,.)$. Hence,
\begin{align}
& V^{t+1}_M(x_{t+1},W_{t+1};\theta) - V^{t}_M(x_{t},W_{t};\theta) = \sum_{k = 0}^{j} c_{t+1+k}(\phi^{t+1}(k, x_{t+1}, \text{M}^{t+1}_p, W_{t+1}), \text{M}^{t+1}_p[k+1]) \nonumber \\
& + \sum_{k = j+1}^{M-1}c_{t+1+k}(\phi^{t+1}(k, x_{t+1}, \text{M}^{t+1}_p, W_{t+1}), \text{M}^{t+1}_p[k+1]) \nonumber \\
& - \sum_{k = 0}^{M-1} c_{t+k}(\phi^t(k, x_t, \text{M}^t_p, W_t), \text{M}^t_p[k+1]) \nonumber \\
& \leq \sum_{k = 0}^{j-1} c_{t+1+k}(\phi^{t+1}(k, x_{t+1}, \tilde{\text{M}}^{t+1}_p, W_{t+1}), \tilde{\text{M}}^{t+1}_p[k+1]) \nonumber \\
& + \sum_{k = j}^{M-1}c_{t+1+k}(\phi^{t+1}(k, x_{t+1},\tilde{\text{M}}^{t+1}_p, W_{t+1}), \tilde{\text{M}}^{t+1}_p[k+1]) \nonumber \\
& - \sum_{k = 0}^{M-1} c_{t+k}(\phi^t(k, x_t, \text{M}^t_p, W_t), \text{M}^t_p[k+1]), \nonumber
\end{align}

where $\tilde{\text{M}}^{t+1}_p = [\text{M}^{t}_p[2:j+1], \text{M}^{*j,t+1}_p[j+1:M]]$, where $\text{M}^{*j,t+1}_p[j+1:M]$ is the optimal control sequence given the control for the first $j$ time steps is $\text{M}^{t}_p[2:j+1]$. The second inequality follows from the fact that $\tilde{\text{M}}^{t+1}_p$ sub optimal compared to $\text{M}^{t+1}_p$. Now by definition
\begin{align}
& \sum_{k = 0}^{M-1} c_{t+k}(\phi^t(k, x_t, \text{M}^t_p, W_t), \text{M}^t_p[k+1]) = c_t(x_t,\text{M}^t_p[1])  \nonumber \\
& + \sum_{k = 0}^{M-2} c_{t+1+k}(\phi^{t+1}(k, f_t(x_t,\text{M}^t_p[1],w_t;\theta), \text{M}^t_p[2:M], W_{t+1}), \text{M}^t_p[k+2]). \nonumber  
\end{align}

Let $\tilde{x}_{t+1} = f_t(x_t,\text{M}^t_p[1],w_t;\theta)$. Then $\tilde{x}_{t+1} = f_t(x_t,\kappa_M(x_t;\theta),w_t;\theta)$. Recall that $x_{t+1} = f_t(x_t,\kappa_M(x_t;\hat{\theta}),w_t;\theta)$. Hence, by Lemma \ref{lem:Lipschitz}
\begin{equation}
x_{t+1} - \tilde{x}_{t+1} = f_t(x_t,\kappa_M(x_t;\hat{\theta}),w_t;\theta) - f_t(x_t,\text{M}^t_p[1],w_t;\theta) \leq \alpha_f \alpha_\kappa S \norm{\hat{\theta}-\theta}_2. \nonumber 
\end{equation}

Hence,
\begin{equation}
\phi^{t+1}(k, \tilde{x}_{t+1}, \text{M}^{t+1}_p, W_{t+1}) - \phi^{t+1}(k, x_{t+1}, \text{M}^{t+1}_p, W_{t+1}) \leq \alpha^{k+1}_f \alpha_\kappa S^{k+1} \norm{\hat{\theta}-\theta}_2 \leq \tilde{\alpha}_f \alpha_\kappa \norm{\hat{\theta}-\theta}_2, \nonumber 
\end{equation}
where $\tilde{\alpha}_f = \max_{k \in [0,1,...,M-2]} \alpha^{k+1}_fS^{k+1}$. Hence,
\begin{align}
& -\sum_{k = 0}^{M-1} c_{t+k}(\phi^t(k, x_t, \text{M}^t_p, W_t), \text{M}^t_p[k+1]) \leq - c_t(x_t,\text{M}^t_p[1])  \nonumber \\
& - \sum_{k = 0}^{M-2} c_{t+1+k}(\phi^{t+1}(k, x_{t+1}, \text{M}^t_p[2:M], W_{t+1}), \text{M}^t_p[k+2]) + \alpha_c\tilde{\alpha}_f \alpha_\kappa(M-1)\norm{\hat{\theta}-\theta}_2. \nonumber  
\end{align}

Hence,
\begin{align}
& V^{t+1}_M(x_{t+1},W_{t+1};\theta) - V^{t}_M(x_{t},W_{t};\theta) \leq \sum_{k = 0}^{j-1} c_{t+1+k}(\phi^{t+1}(k, x_{t+1}, \tilde{\text{M}}^{t+1}_p, W_{t+1}), \tilde{\text{M}}^{t+1}_p[k+1]) \nonumber \\
& + \sum_{k = j}^{M-1}c_{t+1+k}(\phi^{t+1}(k, x_{t+1}, \tilde{\text{M}}^{t+1}_p, W_{t+1}), \tilde{\text{M}}^{t+1}_p[k+1]) - c_t(x_t,\text{M}^t_p[1])  \nonumber \\
& - \sum_{k = 0}^{M-2} c_{t+1+k}(\phi^{t+1}(k, x_{t+1}, \text{M}^t_p[2:M], W_{t+1}), \text{M}^t_p[k+2]) + \alpha_c\tilde{\alpha}_f \alpha_\kappa(M-1)\norm{\hat{\theta}-\theta}_2. \nonumber
\end{align}

Hence,
\begin{align}
& V^{t+1}_M(x_{t+1},W_{t+1};\theta) - V^{t}_M(x_{t},W_{t};\theta) \leq \sum_{k = j}^{M-1}c_{t+1+k}(\phi^{t+1}(k, x_{t+1}, \tilde{\text{M}}^{t+1}_p, W_{t+1}), \tilde{\text{M}}^{t+1}_p[k+1]) \nonumber \\
& - c_t(x_t,\text{M}^t_p[1]) + \alpha_c\tilde{\alpha}_f \alpha_\kappa(M-1)\norm{\hat{\theta}-\theta}_2. \nonumber
\end{align}

By definition of $\tilde{\text{M}}^{t+1}_p$, the first sum on the right is the optimal cost-to-go starting from the state $\phi^{t+1}(j, x_{t+1}, \tilde{\text{M}}^{t+1}_p, W_{t+1})$ for the horizon $M-j$ from $t+j$. Hence, 
\begin{align}
& V^{t+1}_M(x_{t+1},W_{t+1};\theta) - V^{t}_M(x_{t},W_{t};\theta) \leq V^{t+j}_{M-j}(\phi^{t+1}(j, x_{t+1}, \tilde{\text{M}}^{t+1}_p, W_{t+1}), W_{t+j}) \nonumber \\
& - c_t(x_t,\text{M}^t_p[1]) + \alpha_c\tilde{\alpha}_f \alpha_\kappa(M-2)\norm{\hat{\theta}-\theta}_2. \nonumber 
\end{align}

By the assumption in the Lemma we have that
\begin{equation}
V^{t+j}_{M-j}(\phi^{t+1}(j, x_{t+1}, \text{M}^{t+1}_p, W_{t+1}), W_{t+j};\theta) \leq \overline{\alpha} \sigma(\phi^{t+1}(j, x_{t+1}, \tilde{\text{M}}^{t+1}_p, W_{t+1})) + \overline{\gamma} \sum_{k=t+j}^{t+M-2} \norm{w_k}^2_2. \nonumber
\end{equation}

Now, $\phi^{t+1}(j, x_{t+1}, \tilde{\text{M}}^{t+1}_p, W_{t+1}) = \phi^{t+1}(j, x_{t+1}, \text{M}^{t}_p[2:j+1], W_{t+1})$. Hence, 
\begin{align}
& V^{t+1}_M(x_{t+1},W_{t+1};\theta) - V^{t}_M(x_{t},W_{t};\theta) \leq \overline{\alpha} \sigma(\phi^{t+1}(j, x_{t+1}, \text{M}^{t}_p[2:j+1], W_{t+1})) \nonumber \\
& + \overline{\gamma} \sum_{k=t+j}^{t+M-2} \norm{w_k}^2_2 - c_t(x_t,\text{M}^t_p[1]) + \alpha_c\tilde{\alpha}_f \alpha_\kappa(M-1)\norm{\hat{\theta}-\theta}_2.
\label{eq:lemma-Vdiff-ub-unknownsys}
\end{align}

From Assumption \ref{ass:stagecost} and the fact that $V^t_M(x_t, W_{t:t^e};\theta) \leq \overline{\alpha}\sigma(x_t) + \overline{\gamma} \sum_{k=t}^{t^e-1} \norm{w_t}^2_2$ for any $t$ it follows that
\begin{align}
& \underline{\alpha} \sum_{k = 0}^{M-2} \sigma(\phi^{t+1}(k, x_{t+1}, \text{M}^{t}_p[2:k+1], W_{t+1})) \nonumber \\
& \leq \sum_{k = 0}^{M-2} c_{t+1+k}(\phi^{t+1}(k, x_{t+1}, \text{M}^{t}_p[2:k+1], W_{t+1}), \text{M}^{t}_p[k+2]) \nonumber \\
& \leq c_t(x_t,\text{M}^{t}_p[1])+ \sum_{k = 0}^{M-2} c_{t+1+k}(\phi^{t+1}(k, x_{t+1}, \text{M}^{t+1}_p, W_{t+1}), \text{M}^{t+1}_p[k+1]) \nonumber \\
& \leq c_t(x_t,\text{M}^{t}_p[1])+ \sum_{k = 0}^{M-2} c_{t+1+k}(\phi^{t+1}(k, \tilde{x}_{t+1}, \text{M}^t_p[2:M], W_{t+1}), \text{M}^{t}_p[k+2]) \nonumber \\
& + \alpha_c\tilde{\alpha}_f \alpha_\kappa(M-1)\norm{\hat{\theta}-\theta}_2 \leq \overline{\alpha}\sigma(x_{t}) + \overline{\gamma} \sum_{k=t}^{t^e} \norm{w_k}^2_2 + \alpha_c\tilde{\alpha}_f \alpha_\kappa(M-1)\norm{\hat{\theta}-\theta}_2. \nonumber 
\end{align}

The above inequality implies that $\exists \ j^{*} \in \{0,1,...,M-2\}$ such that
\begin{align}
& \sigma(\phi^{t+1}(j^{*}, x_{t+1}, \text{M}^{t}_p[2:j^{*}+1], W_{t+1})) \nonumber \\
& \leq \frac{1}{\underline{\alpha}(M-1)}\left(\overline{\alpha}\sigma(x_{t}) + \overline{\gamma} \sum_{k=t}^{t^e} \norm{w_k}^2_2  + \alpha_c\tilde{\alpha}_f \alpha_\kappa(M-1)\norm{\hat{\theta}-\theta}_2\right). \nonumber 
\end{align}

Then using this in Eq. \eqref{eq:lemma-Vdiff-ub-unknownsys} it we get that
\begin{align}
& V^{t+1}_M(x_{t+1},W_{t+1};\theta) - V^{t}_M(x_{t},W_{t};\theta) \leq \overline{\alpha} \sigma(\phi^{t+1}(j^{*}, x_{t+1}, \text{M}^{t}_p[2:j^{*}+1], W_{t+1})) \nonumber \\
& + \overline{\gamma} \sum_{k=t+j^{*}}^{t+M-2} \norm{w_k}^2_2 - c_t(x_t,\text{M}^t_p[1]) + \alpha_c\alpha_f \alpha_k(M-1)\norm{\hat{\theta}-\theta}_2 \nonumber \\
& \leq \frac{\overline{\alpha}}{\underline{\alpha}(M-1)}\left(\overline{\alpha}\sigma(x_{t}) + \overline{\gamma} \sum_{k=t}^{t^e} \norm{w_k}^2_2  + \alpha_c\tilde{\alpha}_f \alpha_\kappa(M-1)\norm{\hat{\theta}-\theta}_2\right) + \overline{\gamma} \sum_{k=t+j^{*}}^{t+M-2} \norm{w_k}^2_2 \nonumber \\
& - c_t(x_t,\text{M}^t_p[1]) + \alpha_c\tilde{\alpha}_f \alpha_\kappa(M-1)\norm{\hat{\theta}-\theta}_2. \nonumber
\end{align}

Then using the fact that $c_t(x_t,\text{M}^t_p[1]) \geq \underline{\alpha}\sigma(x_t)$ and rearranging terms we get that
\begin{align}
& V^{t+1}_M(x_{t+1},W_{t+1};\theta) - V^{t}_M(x_{t},W_{t};\theta) \leq \frac{\overline{\alpha}^2}{\underline{\alpha}(M-1)}\sigma(x_{t}) - \underline{\alpha}\sigma(x_t) \nonumber \\
& + \overline{\gamma}\left(\frac{\overline{\alpha}}{\underline{\alpha}(M-1)}+1\right)\left(\sum_{k=t}^{t^e} \norm{w_k}^2_2\right) + \alpha_c\tilde{\alpha}_f \alpha_\kappa(M-2)\left(\frac{\overline{\alpha}}{\underline{\alpha}(M-1)}+1\right)\norm{\hat{\theta}-\theta}_2\nonumber \\
& = \left(\frac{\overline{\alpha}^2}{\underline{\alpha}(M-1)}-\underline{\alpha}\right)\sigma(x_{t}) + \overline{\gamma}\left(\frac{\overline{\alpha}}{\underline{\alpha}(M-1)}+1\right)\left(\sum_{k=t}^{t^e} \norm{w_k}^2_2\right) \nonumber \\
& + \alpha_c\tilde{\alpha}_f \alpha_\kappa(M-1)\left(\frac{\overline{\alpha}}{\underline{\alpha}(M-1)}+1\right)\norm{\hat{\theta}-\theta}_2. \nonumber
\end{align}

Hence,
\begin{align}
& V^{t+1}_M(x_{t+1},W_{t+1};\hat{\theta}) - V^{t}_M(x_{t},W_{t};\hat{\theta}) \leq \left(\frac{\overline{\alpha}^2}{\underline{\alpha}(M-1)}-\underline{\alpha}\right)\sigma(x_{t}) \nonumber \\
& + \overline{\gamma}\left(\frac{\overline{\alpha}}{\underline{\alpha}(M-1)}+1\right)\left(\sum_{k=t}^{t^e} \norm{w_k}^2_2\right) + \left(2\alpha_V+ \alpha_c\tilde{\alpha}_f \alpha_\kappa(M-1)\left(\frac{\overline{\alpha}}{\underline{\alpha}(M-1)}+1\right)\right)\norm{\hat{\theta}-\theta}_2\ \blacksquare \nonumber
\end{align}

\section*{Appendix G: Proof of Lemma \ref{lem:cost-bound-unknownsystem}}

The proof is very similar to Lemma \ref{lem:cost-bound-unknownsystem}. Since $M > \overline{\alpha}^2/\underline{\alpha}^2+1$, there exists $\tilde{\epsilon} \in (0,\underline{\alpha})$ such that 
\begin{equation}
M > \overline{\alpha}^2/\left(\underline{\alpha}(\underline{\alpha}-\tilde{\epsilon})\right) + 1 \geq \overline{\alpha}/\left(\underline{\alpha}-\tilde{\epsilon}\right)+1. \nonumber
\end{equation}

For convenience of presentation in the following we denote $V^{t}_M(x_{t},W_{t:t^e};\hat{\theta})$ by $V^{t}_M$. From Lemma \ref{lem:Lyapunov-existence} it follows that
\begin{align}
& V^{t+1}_M - V^{t}_M \leq \Gamma_V \sigma(x) + \Gamma^{\gamma}_V \sum_{k=t}^{t^e} \norm{w_k}^2_2 + \Gamma^\theta_V \norm{\hat{\theta}-\theta}_2, \nonumber \\
& \Gamma_V = \left(\frac{\overline{\alpha}^2}{\underline{\alpha}(M-1)}-\underline{\alpha}\right), \ \Gamma^{\gamma}_V = \overline{\gamma}\left(\frac{\overline{\alpha}}{\underline{\alpha}(M-1)}+1\right), \nonumber \\
& \Gamma^\theta_V = \left(2\alpha_V+ \alpha_c\tilde{\alpha}_f \alpha_\kappa(M-1)\left(\frac{\overline{\alpha}}{\underline{\alpha}(M-1)}+1\right)\right). \nonumber
\end{align}

From the above inequality it follows that $\Gamma_V \leq -\tilde{\epsilon}$ and $\Gamma^{\gamma}_V \leq \overline{\gamma}\left((\underline{\alpha}-\tilde{\epsilon})/\overline{\alpha} + 1\right)$. Hence,
\begin{align}
& V^{t+1}_M - V^{t}_M \leq -\tilde{\epsilon} \sigma(x) + \overline{\gamma}\left(\frac{\underline{\alpha}-\tilde{\epsilon}}{\overline{\alpha}} + 1\right) \sum_{k=t}^{t^e} \norm{w_k}^2_2 \nonumber \\
& + \left(2\alpha_V+ \alpha_c\tilde{\alpha}_f \alpha_\kappa(M-1)\left(\frac{\underline{\alpha}-\tilde{\epsilon}}{\overline{\alpha}} + 1\right)\right) \norm{\hat{\theta}-\theta}_2. \nonumber
\end{align}

Then 
\begin{equation}
V^t_M \leq V^t_M(x_t,W_{t:t^e};\theta) + \alpha_V\norm{\hat{\theta}-\theta}_2 \leq \overline{\alpha}\sigma(x_t) + \overline{\gamma} \sum_{k=t}^{t^e-1} \norm{w_t}^2_2 + \alpha_V\norm{\hat{\theta}-\theta}_2. \nonumber 
\end{equation}

Hence, 
\begin{align}
V^{t+1}_M - V^{t}_M & \leq -\frac{\tilde{\epsilon}}{\overline{\alpha}}\left( V^t_M - \overline{\gamma} \sum_{k=t}^{t^e-1} \norm{w_k}^2_2- \alpha_V\norm{\hat{\theta}-\theta}_2\right) + \overline{\gamma}\left(\frac{\underline{\alpha}-\tilde{\epsilon}}{\overline{\alpha}} + 1\right) \sum_{k=t}^{t^e} \norm{w_k}^2_2 \nonumber \\
& + \left(2\alpha_V+ \alpha_c\tilde{\alpha}_f \alpha_\kappa(M-1)\left(\frac{\underline{\alpha}-\tilde{\epsilon}}{\overline{\alpha}} + 1\right)\right) \norm{\hat{\theta}-\theta}_2 \nonumber\\
& = -\frac{\tilde{\epsilon}}{\overline{\alpha}} V^t_M + \overline{\gamma}\left(\frac{\underline{\alpha}}{\overline{\alpha}} + 1\right) \sum_{k=t}^{t^e} \norm{w_k}^2_2 \nonumber \\
& + \left(\alpha_V\left(2+\frac{\tilde{\epsilon}}{\overline{\alpha}}\right)+ \alpha_c\tilde{\alpha}_f \alpha_\kappa(M-1)\left(\frac{\underline{\alpha}-\tilde{\epsilon}}{\overline{\alpha}} + 1\right)\right) \norm{\hat{\theta}-\theta}_2. \nonumber
\end{align}

Pick $\epsilon$ such that $0 < \epsilon \leq \tilde{\epsilon}/\overline{\alpha}$. Then
\begin{align}
& V^{t+1}_M - V^{t}_M \leq - \epsilon V^t_M + \overline{\gamma}\left(\frac{\underline{\alpha}}{\overline{\alpha}} + 1\right) \sum_{k=t}^{t^e} \norm{w_k}^2_2 \nonumber \\
& + \left(\alpha_V\left(2+\frac{\tilde{\epsilon}}{\overline{\alpha}}\right)+ \alpha_c\tilde{\alpha}_f \alpha_\kappa(M-1)\left(\frac{\underline{\alpha}-\tilde{\epsilon}}{\overline{\alpha}} + 1\right)\right) \norm{\hat{\theta}-\theta}_2. \nonumber
\end{align}

Let $a = (1- \epsilon)$, where $a < 1$. Let $c = \left(\alpha_V\left(2+\frac{\tilde{\epsilon}}{\overline{\alpha}}\right)+ \alpha_c\tilde{\alpha}_f \alpha_\kappa(M-1)\left(\frac{\underline{\alpha}-\tilde{\epsilon}}{\overline{\alpha}} + 1\right)\right)$, and $b = \overline{\gamma}\left(\frac{\underline{\alpha}}{\overline{\alpha}} + 1\right)$. Then
\begin{equation}
V^{t+1}_M \leq  a V^t_M + b \sum_{k=t}^{t^e} \norm{w_k}^2_2 + c \norm{\hat{\theta}-\theta}_2. \nonumber
\end{equation}

Then following steps similar to the proof of Lemma \ref{lem:cost-bound-unknownsystem} we get that 
\begin{align}
& V^{t+H}_M \leq a^H V^{t}_M + \frac{ba^{H-M}}{1-a}\left(\sum_{k=0}^{M-1}\norm{w_{t+k}}^2_2\right) + \frac{b}{1-a}\sum_{k=M}^{H-1}a^{H-k-1}\norm{w_{t+k}}^2_2 \nonumber \\
& + \frac{b}{1-a}\sum_{k = H}^{H+M-2} \norm{w_{t+k}}^2_2 + c\frac{1-a^H}{1-a}\norm{\hat{\theta}-\theta}_2. \nonumber 
\end{align}

Since $V^{t+H}_M \geq c_{t_H}(x_{t_H},\kappa_M(x_{t_H};\hat{\theta}))$, the condition $V^t_M \leq \overline{\alpha}\sigma(x_t) + \overline{\gamma} \sum_{k=t}^{t^e-1} \norm{w_t}^2_2 + \alpha_V\norm{\hat{\theta}-\theta}_2$ implies
\begin{align}
& V^{t+H}_M \leq a^H \overline{\alpha} \sigma(x_t) + \left(\frac{ba^{H-M}}{1-a} + a^H\overline{\gamma}\right)\left(\sum_{k=0}^{M-1}\norm{w_{t+k}}^2_2\right) + \frac{b}{1-a}\sum_{k=M}^{H-1}a^{H-k-1}\norm{w_{t+k}}^2_2 \nonumber \\
& + \frac{b}{1-a}\sum_{k = H}^{H+M-2} \norm{w_{t+k}}^2_2 + \left( c\frac{1-a^H}{1-a} +a^H \alpha_V\right)\norm{\hat{\theta}-\theta}_2\ \blacksquare \nonumber 
\end{align}

\end{document}